\documentclass[a4paper,11pt]{article}
\pdfoutput=1 

\usepackage{jheppub} 

\usepackage[T1]{fontenc} 
\usepackage{empheq}
\usepackage[table]{xcolor}
\usepackage{hyperref}
\hypersetup{
  colorlinks=true,
  citecolor=blue,
  linkcolor=newgreen,
  filecolor=blue,      
  urlcolor=teal,
}
\usepackage{tikz}

\usetikzlibrary{positioning}
\usetikzlibrary{arrows}
\usetikzlibrary{positioning,decorations.pathmorphing,decorations.markings,arrows}

\usetikzlibrary{math}
\tikzmath{\qq=0.22;}
\definecolor{neworange}{RGB}{255,110,30}
\definecolor{newteal}{RGB}{0,182,176}
\definecolor{newgold}{RGB}{169,108,8}
\definecolor{newgreen}{RGB}{10,127,76}
\definecolor{lightcyan}{RGB}{224,255,255}
\tikzmath{\s=7;} 
\tikzmath{\ror=\qq*0.41421/1.41421;}
\tikzset{
mystyle/.style={
  circle,
  inner sep=0pt,
  text width=7mm,
  align=center,
  draw=black,
  fill=white
  }
}
\tikzset{
	ncbar angle/.initial=90,
	ncbar/.style={
		to path=(\tikztostart)
		-- ($(\tikztostart)!#1!\pgfkeysvalueof{/tikz/ncbar angle}:(\tikztotarget)$)
		-- ($(\tikztotarget)!($(\tikztostart)!#1!\pgfkeysvalueof{/tikz/ncbar angle}:(\tikztotarget)$)!\pgfkeysvalueof{/tikz/ncbar angle}:(\tikztostart)$)
		-- (\tikztotarget)
	},
	ncbar/.default=0.5cm,
}

\tikzset{square left brace/.style={ncbar=0.5cm}}
\tikzset{square right brace/.style={ncbar=-0.5cm}}

\tikzset{round left paren/.style={ncbar=0.5cm,out=120,in=-120}}
\tikzset{round right paren/.style={ncbar=0.5cm,out=60,in=-60}}
\usetikzlibrary{shapes,arrows,cd,chains,decorations.markings,decorations.pathmorphing,calc,positioning,patterns}

\tikzset{
	->-/.style args={#1rotate#2}{decoration={markings, mark=at position #1 with {\arrow[scale=1.5,rotate = #2 ]{stealth}}}, postaction={decorate}}
}
\tikzset{
	-r-/.style args={#1rotate#2}{decoration={markings, mark=at position #1 with {\arrow[scale=1,rotate = #2 ]{>}}}, postaction={decorate}}
}

\usepackage{amsmath}
\usepackage{mathtools}
\usepackage{bbm}
\usepackage{nicematrix}
\usepackage[normalem]{ulem}
\usepackage{tabulary}

\newcommand*\widefbox[1]{\fbox{\hspace{1.2em}#1\hspace{1.2em}}}

\newcolumntype{L}{>{$}l<{$}}
\newcolumntype{C}{>{$}c<{$}}
\definecolor{silver}{rgb}{0.8,0.8,0.8}

\title{\boldmath  Duality defects in $D_n$-type Niemeier lattice CFTs}

\author[a,c]{Sachin Grover}
\author[b,c]{, Subramanya Hegde}
\author[a,c]{, Dileep P. Jatkar}

\affiliation[a]{Harish-Chandra Research Institute, Chhatnag Road, Jhunsi, Allahabad, India 211019. }
\affiliation[b]{The Institute of Mathematical Sciences, IV Cross Road, CIT Campus, Taramani, Chennai, India 600113. }
\affiliation[c]{Homi Bhabha National Institute, Training School Complex, Anushakti Nagar, Mumbai, India 400085. }

\emailAdd{sachingrover@hri.res.in}
\emailAdd{subbuh@imsc.res.in}
\emailAdd{dileep@hri.res.in}

\abstract{We discuss the construction of duality defects in $c=24$ meromorphic CFTs that correspond to Niemeier lattices. We will illustrate our constructions for the $D_n$-type lattices. We will identify non-anomalous $\mathbb{Z}_2$ symmetries of these theories, and we show that on orbifolding with respect to these symmetries, these theories map to each other. We investigate this map, and in the case of self-dual orbifolds, we provide the duality defect partition functions. We show that exchange automorphisms in some CFTs give rise to a new class of defect partition functions.}

\begin{document}
\maketitle
\flushbottom
\allowdisplaybreaks

\newcommand{\bm}[1]{m_{\ell_{#1}}}

\section{Introduction}
\label{sec:introduction}

Two dimensional conformal field theory(CFT)\cite{Belavin:1984vu} has seen renewed interests in recent times.  These developments have happened in the classification problem through the modular bootstrap program\cite{Hampapura:2015cea,Gaberdiel:2016zke,Mukhi:2019xjy,Mukhi:2020sxt,Mukhi:2022bte,Das:2020wsi,Das:2021uvd,Das:2022slz,Das:2022uoe,Das:2023qns,Gowdigere:2023xnm} as well as through the 4D-2D correspondence\cite{Beem:2013sza}.  The program of classification of conformal field theories was initiated in the late 1980s\cite{Anderson:1987ge,Mathur:1988na,Mathur:1989pk} with the hope of classifying rational CFT(RCFT)\cite{Moore:1988qv} with a fixed number of characters.  While some progress was made in classifying two character theories using the modular differential equation method as well as by seeking solutions of the Diophantine equations\cite{Mathur:1988na,Mathur:1989pk}, it was quickly realized that the classification, in general, is a pretty daunting task.  After a long period of relatively low-key activity, the classification problem was resurrected recently and has led to the complete classification of two character RCFTs with $c<25$\cite{Mukhi:2019xjy,Mukhi:2020sxt,Mukhi:2022bte,Rayhaun:2023pgc}.

The 2D CFT continued to stay in limelight because of their relevance to AdS$_{3}$ holography but the CFTs of interest in this case were theories with large central charge\cite{Maldacena:2000hw,Maldacena:2000kv,Maldacena:2001km,Eberhardt:2018ouy,Eberhardt:2019niq,Eberhardt:2019qcl,Eberhardt:2019ywk,Gaberdiel:2020ycd,Gaberdiel:2021jrv,Gaberdiel:2021kkp,Gaberdiel:2021qbb,Ashok:2021iqn,Ashok:2022thd,Gaberdiel:2022oeu,Gaberdiel:2023lco,Ashok:2023kkd}, quite different from those pursued in the classification problem.  Study of different correlation functions, like those involving two heavy and two light operators shed light on the dual relationship between the CFT and the bulk gravity theory.  The AdS$_{3}$ holography also led to a proposal of the pure gravity in AdS$_{3}$ described in terms of meromorphic CFT with a central charge $c=24n$, with large $n$\cite{Witten:2007kt}.  While this proposal had limited success, it brought the attention back to the meromorphic CFTs ({\it i.e.}, CFTs with $c=24n$)\cite{Gaiotto:2007xh,Gaberdiel:2007ve}.  The meromorphic CFTs with $c=24$ will be the focus of our attention in this manuscript.

Another interesting development is triggered by study of topological defects in d-dimensional CFT\cite{Bhardwaj:2017xup,Kaidi:2021gbs,Kaidi:2021xfk,Brennan:2022tyl,Bhardwaj:2022kot,Kaidi:2022cpf,Kaidi:2022sng,Bhardwaj:2023ayw,Bhardwaj:2023wzd,Kaidi:2023maf,Schafer-Nameki:2023jdn}.  The defects are extended states in the CFT, generated by line operators or surface operators, etc.  They dig out new symmetries in these theories.  In particular, they are related to higher homotopy groups, which unlike the fundamental group, are abelian and as a result associated higher groups are abelian symmetries which characterize these defects in the CFT.  These symmetries associated with defects are intimately connected with the 't Hooft anomalies and shed light on the behaviour of perturbed CFTs under the renormalization group transformations\cite{Brennan:2022tyl}.  These higher groups also define the generalized fusion categories for defect Hilbert spaces.  In general, these fusion categories also include non-invertible defects.  In 2D CFT, one can have topological defect lines which can be invertible or non-invertible. They provide insights into the topological structure of the theory\cite{Chang:2018iay}.  The defining property that these topological defects operators possess is that they commute with the Virasoro algebra.  These operators satisfy the fusion algebra and are associated with higher group symmetries.  Using these topological line operators, one can define twisted Hilbert spaces, and the twisted partition function obtained from this Hilbert space has to possess positive semi-definite $q$ series expansion, ensuring that it can be interpreted as the degeneracy of states in the twisted sector\cite{Chang:2018iay,Lin:2019hks,Lin:2019kpn,Lin:2021udi,Hegde:2021sdm,Burbano:2021loy}.  This imposes a strong constraint on the topological line operators and restricts the number of consistent topological line operators.

The meromorphic CFTs\cite{Goddard:1989dp} with central charge $c=24$ are interesting in their own right.  These CFTs possess holomorphic factorization mainly due to the fact that all the Virasoro primaries in these theories have integral conformal dimensions\cite {Witten:1991mm}.  As a result, the entire chiral partition function is combined into a single module.   The classification of 2D CFT with two characters for c < 25 relies on the classification of meromorphic CFTs for c < 32.  In fact, the single character theories occur at a central charge multiple of 8, but these theories are not holomorphic due to the non-trivial transformation property under the modular $T$ transformation.  The chiral partition function of $c=24$ CFT transforms trivially under the modular $T$ transformation, and hence it is modular invariant.  This property of the chiral partition function ensures the holomorphic factorization.  Among the single character theories, we have only one candidate at $c=8$, namely the $E_{8,1}$ theory and at $c=16$, we have two candidate theories, namely, $E_{8,1}\times E_{8,1}$ and $SO(32)_{1}$.  The structure becomes quite rich at $c=24$ with Schellekens\cite{Schellekens:1992db} classifying 71 CFTs belonging to this central charge with holomorphic decomposition.  The Niemeier lattice theories belonging to even self-dual lattices corresponding to level 1 Ka\v c-Moody algebra symmetry form a subset of theories classified by Schellekens.  Leaving out that Monster CFT, all other theories have current algebra symmetry and the modular invariant partition function is given by the Hauptmodul $j(\tau)$ up to the addition of an integer $N \ge -744$.  A priori, it appears that any value of $N$ subject to the lower bound can lead to a legitimate meromorphic CFT, but that is not the case, in fact, many allowed cases require non-trivial glue code to obtain the partition function $j(\tau)+N$.  In order to distinguish these theories, it is useful to study topological defects in these theories.  Crucially, one can look for topological defect operators that commute with the Virasoro algebra but not with the full current algebra for the theory.  The structure of these defects and their fusion categories are determined by studying the defect lines in the subgroups obtained by deleting the nodes on the extended Dynkin diagrams or by orbifolding automorphisms of the symmetry groups.

The duality defects have a special status within the topological line defects in CFT because orbifolding this duality symmetry gives back the original theory\cite{Lin:2019hks}.  One classic example of this is the Kramers-Wannier duality in the Ising model.  We will discuss this example in the subsection \ref{subsec:top-def-review}.  We will be interested in studying defect partition functions in the Niemeier lattice CFTs by exploiting automorphisms in the current algebra symmetries of these models.  Although the Niemeier lattices consist of $ADE$ type current algebra lattices, we will focus on theories with the $D$-type current algebra symmetries at level 1.  These examples belong to the Niemeier lattices and the list of CFT with $D$-type current algebra symmetry that we will consider in this manuscript is given by $D_{24,1}$, $D_{20,1}\times D_{4,1}$, $D_{12,1}^{2}$, $D_{8,1}^{3}$, $D_{6,1}^{4}$, and $D_{4,1}^{6}$.

We will review the duality defects in section \ref{sec:review}, where we will also set up our notation and discuss the topological defects, meromorphic CFTs and duality defects in the meromorphic theories.  We will begin with a review of topological defects in 2D CFT and then discuss the duality defect using the example of the Kramers-Wannier duality in the Ising model.  We will then review meromorphic CFTs and the Niemeier lattices, focusing on the D-type lattices, which will be discussed in detail in the next section.  We will then discuss duality defects in the meromorphic CFT using the $\mathbb{Z}_2$ symmetry of the theory.  In section \ref{sec:D-def-key}, we begin our analysis of duality defects in $D_{n}$-type Niemeier lattice CFTs.  In the first example involving $D_{24,1}$ CFT, the extended Dynkin diagram has no outer automorphism consistent with glue code.  We, therefore, look at the inner automorphisms.  For $\mathbb{Z}_{2}$ action, we have two possible ways of using the Ka\v c theorem, either index on one of the nodes with the mark two is non-vanishing or two indices with mark one each is non-vanishing.  We consider the former case (i.e., one index non-vanishing) first, the latter one contains fewer examples and typically lead to non-semisimple groups.  The first case corresponds to breaking $D_{24,1}$ to $D_{r,1}\times D_{24-r,1}$, which decomposes the original lattice of $D_{24,1}$ into an invariant sublattice $L_{0}$ and an orthogonal complement $L-L_{0}$.  The $x$ vector constructed from the weight vector corresponding to the $r$-th node generates twisted sectors, and combining an appropriate choice of the twisted lattice with the invariant sublattice generates another even self-dual lattice.  We demonstrate it for $r=12$, which generates the $(D_{12,1})^{2}$ CFT.  We show that this procedure can be carried out for $D_{8,1}\times D_{16,1}$ and $D_{4,1}\times D_{20,1}$ examples as well.  It turns out that the choice of which twisted lattice to use depends on whether $x\cdot x$ is even or odd.  While for $r=4,12$, it is odd, and the twisted lattice needed for reconstruction of the self-dual lattice is $(L-L_{0})+x$, for $r=8$, it is even, and the twisted lattice is $L_{0}+x$.  We then explicitly construct the duality defects in $D_{4,1}\times D_{20,1}$ and compute the defect partition function.  We then study duality defects in $(D_{12,1})^{2}$, this example has a much richer structure based on different ways of implementing inner automorphisms in two $D_{12,1}$ parts.  This is the first place where we get to implement exchange (outer) automorphism, in which, two subsectors coming from two $D_{12,1}$ are exchanged.  We discuss the $D_{10,1}\times D_{2,1}\times D_{10,1}\times D_{2,1}$ case in detail and show that the exchange $\mathbb{Z}_{2}$ outer automorphism corresponds to a specific exchange of two $A_{1,1}$.  In section \ref{sec:inner-outer-classification}, we tabulate the orbifolds of $D_{24,1}$,  $D_{12,1}^{2}$, $D_{8,1}^{3}$, and $D_{6,1}^{4}$, identify the Lie groups corresponding to each choice of automorphism, and write down the defect partition function.  We conclude with a brief discussion of the application of these results in the discussion section \ref{sec:discussion}.

\section{Review of duality defects in meromorphic CFTs}
\label{sec:review}
To make this article self-contained, we will begin this section with a review of essential aspects of topological defects in two-dimensional meromorphic conformal field theories that are needed to provide context for our results. We will briefly review topological defects in 2D CFTs, emphasising the integrality of the $q$-expansion coefficients in the defect Hilbert space partition function. We will then briefly review meromorphic 2D CFTs and their relation to self-dual lattices. Finally, we will discuss the recently constructed examples of duality defects in these meromorphic CFTs, which will set the stage for our results in the next section.

\subsection{Brief review of topological defects in 2D CFTs}
\label{subsec:top-def-review}
In ordinary quantum mechanics, symmetries are realised as (anti-)unitary operators on rays in the Hilbert space according to Wigner's symmetry representation theorem \cite{wigner2012group,Weinberg:1995mt}. In quantum field theory, one can have more general operators which are non-invertible. Instead of a group structure, now the symmetries satisfy, in general, a fusion category \cite{Chang:2018iay,Shao:2023gho}. Symmetry generators are topological surface operators, and the fusion category describes how they combine. Once the fusion category is known, and the action of these surface operators on the local operators of the theory is known, one has a complete description of the symmetry. These were first found in the context of two dimensional systems \cite{Oshikawa:1996dj,Petkova:2000ip}. The prototypical example is the Kramers-Wannier duality in the Ising model expressed as a non-invertible duality defect \cite{Frohlich:2004ef}.

In 2D CFTs, topological defect operators are line operators which commute with the Virasoro algebra \cite{Petkova:2000ip}. They can run along either a space-like direction or a time-like direction. When they run along a space-like direction, they acquire the meaning of operators in the Hilbert space defined on a space-like slice. When they run along a time-like direction, they twist the boundary condition for each space-like slice they cross, giving rise to a twisted Hilbert space. An important condition that shows that these operators are not generic is the fact that the degeneracies of states at different energies in the twisted Hilbert space are positive integers. One can concretely analyse this condition by considering the partition function with a topological defect running on a space-like slice. In this case, one obtains a partition function with an operator insertion. We can write,
\begin{align}
	\mathcal{Z}^{\eta}=\text{Tr}_H(\hat{\eta} q^{L_0-\frac{c}{24}}\bar{q}^{\bar{L}_0-\frac{\bar{c}}{24}}),
\end{align}
where $H$ is the Hilbert space, $\hat{\eta}$ is the topological operator, $q=e^{2\pi i \tau}$ with $\tau$ being the modular parameter of the torus. Since this is defined on a torus, we can perform a modular transformation to make the topological operator run along the time circle. This gives,
\begin{align}
	\mathcal{Z}_{\eta}(\tau,\bar{\tau})=\mathcal{Z}^{\eta}(-1/\tau,-1/\bar{\tau})=\text{Tr}_{H_{\eta}}( q^{L_0-\frac{c}{24}}\bar{q}^{\bar{L}_0-\frac{\bar{c}}{24}}),
\end{align}
where $H_{\eta}$ is the twisted Hilbert space. Since this is a partition function over the twisted Hilbert space, the $q$-expansion coefficients must be positive integers as they compute the degeneracy. The positivity of the coefficients in the $q$-expansion provides a stringent condition on which topological defects are realised in a given theory. There has been extensive study on these topological defects in two dimensions, especially for rational CFTs. In diagonal rational CFTs, topological defects are classified if we assume that for theories with current algebra symmetry, the topological defects also commute with the current algebra generators \cite{Petkova:2000ip}. In this case, the topological defects are given as Verlinde lines, and the Verlinde formula guarantees the positive integer coefficients for the twisted Hilbert space degeneracies, which are nothing but fusion coefficients \cite{Verlinde:1988sn,Chang:2018iay}. However, note that topological defects that commute with the Virasoro algebra and not with the full current algebra may also exist. It is then interesting to study such operators in the simplest rational CFTs. Memormophic 2D CFTs, which consist of a single current algebra module, provide such examples. Note that as their MTC is trivial, the defect operators one finds can not be used as relevant operators to deform the CFT. Nonetheless, as we have discussed, they contain non-trivial information about the theory as, given a theory, topological defects are not generic operators. In the particular case of duality defect, they imply that the CFT has a self-dual orbifold.

As we will discuss the duality defects in single character c=24 CFTs in this manuscript, it is useful to understand their Ising model realisation. Ising model, at the critical point, maps to a conformal field theory with three primary operators $1, \sigma,\epsilon$ with conformal dimensions $(0,0)$, $(\frac{1}{16},\frac{1}{16})$, $(\frac{1}{2},\frac{1}{2})$. They satisfy the fusion relations,
\begin{align}
	\epsilon \times \epsilon &= 1,\\
	\epsilon \times \sigma &= \sigma \times \epsilon = \sigma,\\
	\sigma \times \sigma &= 1 + \epsilon.
\end{align}
Since the Ising model is a diagonal rational CFT, the topological defects are given by the Verlinde lines. Each Verlinde line corresponds to one of the primaries above, and their fusion rule is isomorphic to the fusion rule of primaries. The topological operators are $1, \eta, \mathcal{N}$ satisfying,
\begin{align}
	\eta \times \eta &= 1,\\
	\eta \times \mathcal{N} &= \mathcal{N} \times \eta = \mathcal{N},\\
	\mathcal{N} \times \mathcal{N} &= 1 + \eta.
\end{align}
We can see from above that the topological defect line (TDL) $\mathcal{N}$ does not contain an inverse under fusion. Therefore, $\mathcal{N}$ is a non-invertible TDL. $\mathcal{N}$ is the duality defect. To see this, note that the TDL $\eta$ is a $\mathbb{Z}_2$ line. It is non-anomalous in the sense that the Ising model can be orbifolded by this $\mathbb{Z}_2$. I.e., the object,
\begin{align}
	\mathcal{Z}_{\text{Orbifold}}&=\frac{1}{2}(\mathcal{Z}+\mathcal{Z}^\eta+\mathcal{Z}_\eta+\mathcal{Z}^\eta_\eta),
\end{align}
is modular invariant. This in particular requires that $T^2\mathcal{Z}_\eta=\mathcal{Z}_\eta$ \cite{Chang:2018iay}. The above orbifold is a self-dual orbifold and $\mathcal{Z}_{\text{Orbifold}}=\mathcal{Z}_{\text{Ising}}$. The defect Hilbert space partition function is given by, 
\begin{align}
	\mathcal{Z}_\eta&=\chi_0(\tau)\bar{\chi}_{\frac{1}{2}}(\bar{\tau})+\chi_{\frac{1}{2}}(\tau)\bar{\chi}_0(\bar{\tau})+\chi_{\frac{1}{16}}(\tau)\bar{\chi}_{\frac{1}{16}}(\bar{\tau}),
\end{align}
where the $\chi_0(\tau),\chi_{\frac{1}{2}}(\tau),\chi_{\frac{1}{16}}(\tau)$ are the chiral Virasoro characters in the modules over the Ising model primaries. We can see above that the coefficients in the $q$ expansion coefficients are positive and indicate the degeneracy in the defect Hilbert space. Generically, defect Hilbert space operators do not have integral values of spin and therefore, lead to branch cuts. However, the defect Hilbert space $H_\eta$ above contains an operator that has conformal dimensions $(\frac{1}{16},\frac{1}{16})$. This operator can be argued to be the disorder operator $\mu$ for the Ising CFT \cite{Frohlich:2004ef,Chang:2018iay}. The self-duality of the Ising model under the $\mathbb{Z}_2$ orbifold is nothing but the Kramers-Wannier duality. However, the TDL referred to as the duality defect is not the $\mathbb{Z}_2$ defect line $\eta$ but the non-invertible defect $\mathcal{N}$. This is because using the fusion rule of $\mathcal{N}$ one can argue that when an $\mathcal{N}$ line passes through the local operator $\sigma$, it converts it to a $\mu$ defect Hilbert space operator which is connected to the $\mathcal{N}$ line with a $\mathcal{N}\mathcal{N}\eta$ $T$-junction. Thus, together with the $\mathbb{Z}_2$ operator $\eta$, the defect operator $\mathcal{N}$ captures the Kramers-Wannier duality of the Ising model. When we study the duality defects in single-character CFTs, it is in this sense of self-dual orbifold that we obtain the duality defect. The TDLs $\eta, \mathcal{N}$ with the fusion rules given above form the simplest examples of the Tambara-Yamagami category where one has one non-invertible element added to a $\mathbb{Z}_N$ fusion rule as \cite{tambara1998tensor},
\begin{align}
	\eta_i \times \eta_j &= \eta_k,\\
	\eta_i \times \mathcal{N} &= \mathcal{N} \times \eta_i = \mathcal{N},\\
	\mathcal{N} \times \mathcal{N} &= 1 + \sum_i \eta_i,
\end{align}
where $\eta_i$ are the $\mathbb{Z}_n$ defect lines. Studying duality defects in rational CFTs is also interesting from the perspective of categorical symmetries. It will be interesting to understand the fusion categories realised in these theories, which provides an additional structure to these theories, even though they have a trivial modular tensor category. When we construct $\mathbb{Z}_2$ duality defects for such theories, the defect Hilbert space partition function also contains information about the fusion kernel appropriate for these defect operators. Fusion kernels for $\mathbb{Z}_2$ Tambara-Yamagami category have been solved while solving for fusion kernels of primaries in \cite{Moore:1988qv}. For the $\mathbb{Z}_2$ Tambara-Yamagami fusion rule, the kernel is determined up to a factor $\epsilon=\pm 1$ as emphasized in equation (6.2) of \cite{Chang:2018iay}. The fusion rule also gives rise to a spin-selection rule for the operators in the defect Hilbert space $H_\mathcal{N}$ according to equation (6.5) of \cite{Chang:2018iay}. Namely,
\begin{align}
	s &\in \pm \frac{1}{16}+\frac{\mathbb{Z}}{2}, \qquad \text{if} \quad \epsilon=1,\label{epsilon_1}\\
	&\in \pm \frac{3}{16}+\frac{\mathbb{Z}}{2}, \qquad \text{if} \quad \epsilon=-1,\label{epsilon_2}
\end{align}
where $s=h-\bar{h}$ is the spin, and for chiral theories to be considered in this manuscript, $\bar{h}=0$ and $s=h$. When $\epsilon=1$, the duality defect is of the Ising category and when $\epsilon=-1$, the duality defect is of the same category as $\text{Ising}\times SU(2)_1 \text{WZW model}$. Crucially, when we calculate the defect Hilbert space partition function $\mathcal{Z}_\mathcal{N}$, we can read off the value of $\epsilon$ from the conformal dimensions and categorise our defect in one of the two categories. This also allows us to use the solution for the fusion kernel and determine the one appropriate for each defect. We find examples of both the categories realised in the meromorphic CFTs we study.

\subsection{Meromorphic 2D CFTs and self dual lattices}
\label{subsec:mer-CFT-review}
Two-dimensional CFTs with rational values for central charge and conformal dimensions, known as rational CFTs (RCFT), are extensively studied \cite{Moore:1988qv,Mukhi:2019xjy}. They display the nice property of holomorphic factorisation \cite{Gaberdiel:2016zke}, which lets one focus on chiral vertex operator algebra modules. For central charges $c=8k$ with $k$ being a positive integer, one can have theories that consist of a single chiral primary, the identity operator. Their MTC is, hence, trivial. Further, if the central charge is $c=24k$, then the theory is consistent as a chiral CFT \cite{Goddard:1989dp}. Theories of such type can be mapped to even self-dual lattices, and they also play a role in lattice compactifications of heterotic string theory \cite{Lerche:1988np}. For $c=24$, their chiral partition function takes the form,
\begin{align}\label{j-function-character}
	\mathcal{Z}(\tau)=j(\tau)-744+N,
\end{align}
where $j(\tau)$ is the modular $j$-function, and $N$ is a positive integer. Schellekens has classified 71 possible values of $N$, corresponding to different meromorphic CFTs \cite{Schellekens:1992db}. One among these is the Monster CFT, while the rest contain a dimension one current algebra and correspond to Lie algebra lattices, which are also listed by Schellekens. Out of these, 23 correspond to Niemeier lattices \cite{niemeier1973definite}. 

The single-character theories discussed above, with central charges $c=8k$, play an important role in the classification of rational CFTs employing modular linear differential equations (MLDE) initiated by Mathur, Mukhi and Sen (MMS) \cite{Mathur:1988na}. The first example at $c=8$ is the $E_{8,1}$ CFT, the unique single-character theory at $c=8$, and it admits commutant pairs of two-character theories as cosets. The list of these commutant pairs is now known as the \emph{MMS series.} While MMS studied the classification of vector-valued modular forms which could act as candidate characters for RCFTs, recently Mukhi and Rayhaun have classified rational CFTs below $c=25$, and their classification relies on the classification of meromorphic CFTs till $c=32$ \cite{Mukhi:2022bte}. The meromorphic CFTs here act as seeds of cosets, where given a two-character theory, one can obtain another two-character theory by taking a coset from a single-character meromorphic CFT. 

In this work, we will construct duality defects for single-character CFTs with $c=24$. Our work is preceded by works which construct topological defects in single character CFTs with $c=8$ \cite{Burbano:2021loy,Hegde:2021sdm} and $c=24$ \cite{Lin:2019hks}. Therefore, reviewing these theories, particularly the $c=24$ Schellekens list CFTs, which are of interest to us, and their relation with Euclidean even self-dual lattices, is useful. To see this relation, we will need to consider vertex operators, which are built out of chiral sigma model fields. Consider fields $X^i$ where $i=1,\ldots, c$ where $c$ is the central charge. One can write the mode expansion as,
\begin{align}
	X^i(z)=x_0^i-ip^i \log (z)+ i\sum_{n\in \mathbb{Z}} \frac{1}{n} \alpha_n^i z^{-n},
\end{align}
where $x_0^i$ denote the vevs for the sigma model fields and $\alpha_n^i$ are oscillator modes. The operator $p^i=\alpha_0^i$ and its eigen-vectors are vertex operators $V_\alpha = c_{\alpha}:e^{i\alpha \cdot X}:$. The eigen-values are $\alpha^i$ and $c_\alpha$ are the cocycle operators\cite{Dolan:1989vr}. Modular invariance on the torus implies that the eigenvalues $\alpha^i$ must form an Euclidean even self-dual lattice \cite{Lerche:1988np}. These can be constructed by using Lie algebra lattices.

Consider vertex operators with $\alpha^i$ as the roots of a simply laced Lie algebra, then the fundamental weights $\omega_i$ obey the relation,
\begin{align}
	(\omega_i,\frac{2\alpha_j}{|\alpha_j|^2})=\delta_{i,j}.
\end{align}
If we normalise the roots to have length 2, then the fundamental weights have integer norms with the roots. Thus, the weight lattice will be the dual of the root lattice\footnote{We will fix $|\alpha_i|^2=2$ from now on, which is equivalent to considering simply laced groups.}. The eigenvalue of the $L_0$ operator for the state that corresponds to the vertex operator $V_\alpha$ is given by $h_\alpha=\frac{(\alpha,\alpha)}{2}$. Note that if we take $\alpha$ as the root of a simply laced Lie algebra, then $h_\alpha=1$. Thus, the vertex operators corresponding to the roots have dimension one and act as currents. However, the root lattice is often not self-dual, and one must add other conjugacy classes. The currents of this self-dual lattice satisfy Ka\v c-Moody algebra at level one. The set of conjugacy classes that need to be added to the root lattice to make the lattice self-dual is known as the glue code. For instance, for $D_{24,1}$, one needs to add the spinor conjugacy class $(s)$ to make it the glue code, and therefore $(s)$ is the glue code. The conjugacy classes in the glue code need to close under addition as one needs to form a lattice. Therefore, it is possible to provide only the generators of the glue code from which one can generate the full glue code. Schellekens provides the list of generators of glue code for each case. Here, we will focus on $D$-type Niemeier lattice CFTs. Their algebras and glue code generators\footnote{In the case of $D_{4,1}^6$, we found that we need to add $(c)^6$ in addition to the glue code generators given in \cite{Conway:1988oqe,Schellekens:1992db} such that the modular invariant partition function is reproduced and also obtain the glue code with order $64$.} are given in Table-\ref{Table-D-type-Schellekens}, as well as their value of $N$ for \eqref{j-function-character}.
\begin{table}
\begin{center}
	\begin{tabular}{ |c| c| c| }
		\hline
		N & Algebra & Glue code generators \\ 
		\hline
		1128 & $D_{24,1}$ & $(s)$ \\  
		\hline
		552 & $D_{12,1}^2$ & $(s,v)+(v,s)$   \\
		\hline
		360 & $D_{8,1}^3$ & $(s,v,v)+(v,v,s)+(v,s,v)$   \\
		\hline
		264 & $D_{6,1}^4$ & $(0,c,s,v)+(0,s,v,c)+(0,v,c,s)+(c,0,v,s)$   \\
		& & $+(c,s,0,v)+(c,v,s,0)+(s,0,c,v)+(s,c,v,0)$\\
		& & $+(s,v,0,c)+(v,0,s,c)+(v,c,0,s)+(v,s,c,0)$\\
		\hline
		168 & $D_{4,1}^6$ &  $(s)^6+(c)^6+(0,0,v,c,c,v)+(0,v,c,c,v,0)$\\
		& & $+(0,c,c,v,0,v)+(0,c,v,0,v,c)+(0,v,0,v,c,c)$ \\
		\hline
	\end{tabular}\label{Table-D-type-Schellekens}
\caption{\label{NL_gluecodes} The table lists the gluecode generators for the $D_n$ Niemeier lattice CFTs and the algebra dimension ($N$). The full gluecode is generated from the list of generators by adding the generators to each other.}
\end{center}
\end{table}

One can find the full glue code from the glue code generators above and the fusion rules of $D$-type algebras. To verify that one has the full glue code, one can verify that when one takes the character for the identity module, which corresponds to the root conjugacy class, as well as the characters for the full glue code, one obtains \eqref{j-function-character}. For instance, for $D_{24,1}$,
\begin{align}
	\mathcal{Z}=j-744+1128=\chi_0^{D_{24,1}}+\chi_s^{D_{24,1}},
\end{align}
where $\chi_0^{D_{24,1}},\chi_s^{D_{24,1}}$ are the characters for the modules with primaries in $(0)_{D_{24}}$ and $(s)_{D_{24}}$ conjugacy classes respectively. Thus, the glue code needed to form a self-dual lattice also reproduces a modular invariant partition function as expected.

\subsection{Duality defects in meromorphic CFTs}
\label{subsec:Def-mer-review}
We will now review the construction of duality defects in single-character CFTs. As mentioned earlier, if one demands that defects commute with the current algebra for the theory, then single-character rational CFTs contain only the trivial identity topological defect line. Their modular tensor category is trivial. However, as we emphasised in section \ref{subsec:top-def-review}, one can define duality defects in the sense of the theory having a self-dual orbifold. Also, defects that commute with the Virasoro algebra and not with the full current algebra can be found, which have positive integer degeneracies in their defect Hilbert space when one takes them to run along the time direction. Such duality defects were first written for the $c=24$ Monster CFT in \cite{Lin:2019hks}. The construction of duality defects in Monster could be viewed in two ways. Monster has two non-anamolous $\mathbb{Z}_2$ automorphisms, which are named $\mathbb{Z}_{2A}$ and $\mathbb{Z}_{2B}$. Orbifolding with respect to $\mathbb{Z}_{2B}$ gives the Leech lattice CFT, whereas Monster is self-dual under the $\mathbb{Z}_{2A}$ orbifold. The latter implies that Monster CFT contains a duality defect line, which forms a Tambara-Yamagami category with the $\mathbb{Z}_{2A}$ line. To find this duality defect line, the authors of \cite{Lin:2019hks} fermionised the Monster CFT with the $\mathbb{Z}_{2A}$ line and the fermionised theory corresponded to a Majorana-Weyl CFT and the fermionic Baby-Monster CFT. The chiral fermion parity in the Majorana-Weyl CFT acted as the duality defect of the Monster CFT, with a multiplication of $\sqrt{2}$ to make the defect Hilbert space consistent by having positive integer degeneracies. Alternatively, the duality defect was also looked at from the perspective of a replacement rule. The partition function of the Monster CFT can be written as,
\begin{align}
	\mathcal{Z}^{\text{Monster}}(\tau)=\chi_0^{\text{Ising}}(\tau)\chi_0^{\text{BM}}(\tau)+\chi_{\frac{1}{2}}^{\text{Ising}}(\tau)\chi_{\frac{3}{2}}^{\text{BM}}(\tau)+\chi_{\frac{1}{16}}^{\text{Ising}}(\tau)\chi_{\frac{31}{16}}^{\text{BM}}(\tau),
\end{align}
where $\chi_h^{\text{Ising}}$ and $\chi_{\tilde{h}}^{\text{BM}}$ are the characters of Ising and Baby-Monster CFTs for modules with primaries of conformal dimensions $h$ and $\tilde{h}$ respectively. Note that in the above, the characters pair when $h+\tilde{h}=2$.  

Contrast the partition function above, with the partition function for the Ising CFT,
\begin{align}
	\mathcal{Z}^{\text{Ising}}(\tau,\overline{\tau})=\chi_0^{\text{Ising}}(\tau)\overline{\chi}_0^{\text{Ising}}(\overline{\tau})+\chi_{\frac{1}{2}}^{\text{Ising}}(\tau)\overline{\chi}_{\frac{1}{2}}^{\text{Ising}}(\overline{\tau})+\chi_{\frac{1}{16}}^{\text{Ising}}(\tau)\chi_{\frac{1}{16}}^{\text{Ising}}(\overline{\tau}).
\end{align}
The duality defect insertion partition function can be obtained from the Verlinde line corresponding to the operator $\sigma$, and reads,
\begin{align}
	\mathcal{Z}^{\text{Ising}\; \mathcal{N}}(\tau,\overline{\tau})=\sqrt{2}\chi_0^{\text{Ising}}(\tau)\overline{\chi}_0^{\text{Ising}}(\overline{\tau})-\sqrt{2}\chi_{\frac{1}{2}}^{\text{Ising}}(\tau)\overline{\chi}_{\frac{1}{2}}^{\text{Ising}}(\overline{\tau}).
\end{align}
If one replaces the anti-chiral Ising characters above with Baby Monster characters, one obtains the duality defect for the Monster CFT corresponding to chiral fermion Majarona-Weyl parity in the fermionised Monster. By performing the $S$-modular transformation, one can get the defect Hilbert space partition function on $H_\mathcal{N}$ and the operators on $H_\mathcal{N}$ were noticed to have the conformal dimensions according to the spin-selection rule derived in \cite{Chang:2018iay} with $\epsilon=1$ in the fusion kernel. Therefore, this is a duality defect in the Ising category. Note that the Baby-Monster CFT and the Ising CFT combine to produce Monster CFT according to the novel coset construction \cite{Mukhi:2020sxt}, with $c+\tilde{c}=24$ and $h+\tilde{h}=2$ when the characters pair up in the partition function analogous to the cases in \cite{Gaberdiel:2016zke}. For the MMS series, one similarly has coset construction of commutant pairs with $c+\tilde{c}=8$, and $h+\tilde{h}=1$. One can then ask if one can realise the Verlinde lines of the two-character theories in the MMS series \cite{Mathur:1988na} and in \cite{Gaberdiel:2016zke} by using the character replacement rule. This question was addressed in \cite{Hegde:2021sdm}, where an interpretation was provided in terms of what part of the current algebra is preserved by these lines with the help of branching rules. In \cite{Burbano:2021loy}, a systematic method was provided to construct $\mathcal{Z}_N$ duality defects in $E_{8,1}$ CFT. In this manuscript, we use this formalism to construct duality defects in Niemeier lattice CFTs. Therefore, we summarise the formalism below by considering the $E_{8,1}$ CFT.

To find $\mathbb{Z}_N$ Tambara-Yamagami category in $E_{8,1}$ CFT, one needs first to identify the non-anomalous $\mathbb{Z}_N$ lines, which, when gauged, give back the same theory. To look for such automorphisms, we need to know $\mathbb{Z}_N$ automorphisms of the $E_{8,1}$ vertex operator algebra. For $E_{8,1}$, there are no outer automorphisms. However, it has inner automorphisms. To diagnose this, one can use Ka\v c-theorem \cite{kac1990infinite}, (a statement can also be found in Theorem 1 in \cite{Burbano:2021loy}). We will discuss the essential elements from Ka\v c-theorem for this manuscript and mention an operational understanding. Level one vertex operator algebras have vertex operators of the type $V_\alpha=c_\alpha :e^{i\alpha\cdot X}:$ as we had mentioned earlier. If one does a translation in the sigma model field $X^i \rightarrow X^i+2\pi x^i$ where $x$ is a known constant vector, then the vertex operator transforms as,
\begin{align}
V_\alpha \rightarrow \zeta_\alpha e^{2\pi i\alpha \cdot x}V_\alpha,
\end{align}
where $\zeta_\alpha = \pm 1$ is a projective phase which can be non-trivial for spinor and conjugate spinor classes of $D_n$-type vertex operator algebras. If $x^i$ belongs to the dual lattice of $\alpha^i$'s then the phase is trivial up to the projective factor. Otherwise, the above is an inner automorphism as it does not transform the lattice vectors $\alpha^i$. To construct $\mathbb{Z}_N$ inner automorphisms, if $\alpha^i$ are simple roots of a simply laced lie algebra of rank $l$, then one can consider,
\begin{align}
x=\frac{1}{N}\sum_{i=1}^l s_i \omega_i,
\end{align}
which act as $\mathbb{Z}_N$ transformations on such vertex operators since $(\alpha_i,\omega_j)=\delta_{i,j}$. Further, if the automorphism is $\mathbb{Z}_N$ on the highest root as well, then it acts as a $\mathbb{Z}_N$ on the entire root diagram. The highest root is,
\begin{align}
\alpha_{\text{high}}=-\sum_{i=1}^l a_i \alpha_i,
\end{align}
where $a_i$ are the marks of the affine ($k=1$) Dynkin diagram. The condition that $x\cdot \alpha_{\text{high}} \in \frac{\mathbb{Z}}{N}$ can be achieved if we have,
\begin{align}
N=s_0+\sum_{i=1}^l a_i s_i,
\end{align}
with $s_0,s_i$ being non-negative integers. If we demand them to be relatively prime, then we can remove the degeneracy of adding with the dual lattice vectors to $x$. The condition above is precisely what Ka\v c-theorem demands for inner automorphisms. For outer automorphisms, Ka\v c-theorem demands,
\begin{align}
N=k(s_0+\sum_{i=1}^l a_i s_i),
\end{align}
where $k=2,3$ corresponds to twisted Dynkin diagrams whose marks are $a_i$. For $E_{8,1}$ one can find $\mathbb{Z}_2$ inner automorphisms from Ka\v c theorem, which we call $\eta$, and by calculating the partition function with insertion $\left(\frac{1+\eta}{2}\right)$ one can find the invariant sector under the inner automorphism. Some $\eta$ may be anomalous in the sense that we can not orbifold with respect to them. As explained earlier, for non-anamolous $\mathbb{Z}_2$, the combination $\frac{1}{2}(\mathcal{Z}+\mathcal{Z}^\eta+\mathcal{Z}_\eta+\mathcal{Z}^\eta_\eta)$ is modular invariant. In particular $\mathcal{Z}^\eta_\eta=T\mathcal{Z}_\eta$ is unambiguous. For anomalous $\mathbb{Z}_2$, this is not true, and they can not be gauged. Thus, this is a discrete 't Hooft anomaly. Note that whenever one has a non-anamolous $\mathbb{Z}_2$, then the $\frac{1}{2}(\mathcal{Z}+\mathcal{Z}^\eta)$ sector is common between the theory and its $\mathbb{Z}_2$ orbifold. Thus, the invariant lattice $L_0 \subset L$ is a part of the self-dual lattice $L^\prime$, corresponding to the orbifolded chiral CFT. Since $E_8$ is the unique self-dual lattice for $8$ dimensions, when $E_{8,1}$ is orbifolded with a non-anomalous $\mathbb{Z}_2$ (or $\mathbb{Z}_N$) it always leads back to $E_{8,1}$ CFT thus leading to self-duality. This uniqueness is not true for $24$ dimensional lattices to be discussed in the next sections, and Niemeier lattices map to each other. 

Returning to the lattice description,  we need to obtain a new self-dual lattice $L^\prime$, which also gives rise to $E_{8,1}$ CFT. We will consider the $\mathbb{Z}_2$ case for illustration. As discussed earlier, vectors of the invariant lattice $L_0$ are common to the lattice for the original theory $L$ and lattice for the orbifolded theory $L^\prime$. Let us call the remaining part of $L$ $L-L_0$. A vector $\alpha^\prime \in L-L_0$ has the property that $\alpha^\prime\cdot x \in \frac{1}{2}+\mathbb{Z}$. With $\alpha \in L_0$, we have instead $\alpha \cdot x \in \mathbb{Z}$. Since the original lattice is self-dual, $\alpha \cdot \alpha^\prime \in \mathbb{Z}$. Now consider the vector $\alpha^\prime+x \in (L-L_0)+x$. This vector has integer inner products with the vectors in $L_0$. If we further demand that vectors in $(L-L_0)+x$ have integer inner products with respect to each other, then we need $x\cdot x \in \mathbb{Z}$. Further if $x\cdot x\in 2\mathbb{Z}+1$, then the vector $\alpha^\prime+x$ has even norm. Therefore, we can take $L^\prime=L_0 \cup \left((L-L_0)+x\right)$ in such cases. We find such cases when we study inner automorphisms of $D$-type lattices in this manuscript. For the case of $E_{8,1}$ CFT as well, non-anamolous $\mathbb{Z}_2$ has $L_0=(0)_{D_8}$, $L-L_0=(s)_{D_8}$ and $x=\frac{\omega_7}{2}\in (v)_{D_8}$ and has an odd norm. The new self dual lattice is $L^\prime=(0)_{D_8} \cup (c)_{D_8}$ since $(s)_{D_8}+(v)_{D_8}=(c)_{D_8}$. In such cases, one can form the following table by looking at the conjugacy classes in $L_0^*$,
\begin{center}
\begin{tabular}{|c|c|}
	\hline
	$L_0$ & $(L-L_0)+x$ \\
	\hline
	$(L-L_0)$ & $L_0+x$\\
	\hline
\end{tabular}
\end{center}
The first column corresponds to $L$, and the first row corresponds to $L^\prime$. Note that if $x\cdot x \in 2\mathbb{Z}$ instead, then we can take $L^\prime=L_0 \cup (L_0+x)$, and the entries in the second column above would be interchanged. Thus, for non-anomalous $\mathbb{Z}_2$, it is necessary and sufficient that $x\cdot x \in \mathbb{Z}$. $x\cdot x \in \mathbb{Z}$ corresponds to the spin selection rule for non-anomalous $\mathbb{Z}_2$ derived in \cite{Chang:2018iay}.

The lattice vectors in the sector $L^\prime - L_0$ correspond to the $H_\eta$ Hilbert space in analogy with the Ising model. Therefore, to find the duality defect line $\mathcal{N}$, we need a line that switches between the rows and columns of the above table, which is analogous to the $\mathcal{N}$ line moving past the operator $\sigma$ in the Ising model to lead to the disorder operator $\mu$. From the lattice perspective, switching the two axes above while keeping $L_0$ common corresponds to an outer automorphism. If we consider a $\mathbb{Z}_2$ outer automorphism $g$ of $L_0$, then we can write the defect insertion partition function for the duality defect $\mathcal{N}$ in the original theory as $\sqrt{2}g$ in the invariant sector:
\begin{align}
	\mathcal{Z}^\mathcal{N}=\text{Tr}_H\left( q^{L_0-\frac{c}{24}}\mathcal{N}\right)=\text{Tr}_H\left(\left(\frac{1+\eta}{2}\right) q^{L_0-\frac{c}{24}}\sqrt{2}g\right).
\end{align}
This guarantees that,
\begin{align}
	\mathcal{Z}^{\mathcal{N}^2}=\text{Tr}_H\left( q^{L_0-\frac{c}{24}}\mathcal{N}^2\right)=\text{Tr}_H\left(\left(1+\eta\right) q^{L_0-\frac{c}{24}}\right),
\end{align}
which corresponds to the Tambara-Yamgami fusion rule $\mathcal{N}^2=1+\eta$. Therefore, one has to look at a $\mathbb{Z}_2$ outer automorphism of the invariant sector under a non-anomalous $\mathbb{Z}_2$ of the full theory \footnote{Note that there will also be such outer automorphisms of the invariant sub lattice which are also automorphisms of the full lattice. These should be excluded while identifying the duality defects.}. This outer automorphism should also be the one that switches between the two lattice descriptions $L$ and $L^\prime$. In the case of $E_{8,1}$, one needs an outer automorphism of $(0)_{D_8}$ which changes $(s)_{D_8}$ to $(c)_{D_8}$. This is nothing but the $\mathbb{Z}_2$ Dynkin diagram automorphism of the $D_8$ Dynkin diagram where the two antennae are switched. i.e., $\alpha_7 \leftrightarrow \alpha_8$, where $\alpha_7,\alpha_8$ are the simple roots corresponding to the antenna's two ends. The invariant lattice then has integer multiples of $\alpha_7+\alpha_8$, a long root. The invariant lattice is a $C_7$ lattice. This also extends to not just the $\alpha$ vectors but also the sigma model fields. One has the exchange $X^7\leftrightarrow X^8$. This fixes the oscillator part of the partition function, and one now has $\frac{1}{\eta^6(\tau)\eta(2\tau)}$ as the oscillator contribution instead of $\eta^8(\tau)$ due to the exchange automorphism. To know the invariant subalgebra, one also needs to know the action on vertex operators, which can acquire phases without changing the lattice vectors. This is determined by Ka\v c-theorem for the twisted Dynkin diagram with $k=2$. Since the $x$-vector is in the dual lattice of the $C_7$ lattice, one finds characters of $B$-type in the defect insertion partition functions. Finally, the duality defect insertion partition function is obtained to be\footnote{The factor $\sqrt{2}$ has been added compared to the expressions in \cite{Burbano:2021loy} such that we have the correct fusion rule and integrality of the $q$-expansion coefficients in the defect Hilbert space partition function $\mathcal{Z}^{E_{8,1}}_\mathcal{N}$.},
\begin{align}
	\mathcal{Z}^{E_{8,1} \; \mathcal{N}}=\sqrt{2}(\chi_{0}^{so(16-i)}\chi_0^{so(i)}-\chi_{v}^{so(16-i)}\chi_v^{so(i)}).
\end{align}
In the next sections, we will find analogous results for $c=24$ Niemeier lattice CFTs of $D$-type. We will use the algorithm of first looking for a non-anomalous $\mathbb{Z}_2$ and then checking if the orbifolding leads to self-duality, which is non-trivial at $c=24$ as explained earlier. We will then find automorphisms of the invariant sector under non-anamolous $\mathbb{Z}_2$. Conceptually, two new things need to be taken care of: one is the presence of the glue code, and the second is the possibility of exchange automorphisms. We should note here that while this work was in progress, three works appeared simultaneously \cite{BoyleSmith:2023xkd,Rayhaun:2023pgc,Hohn:2023auw} where they study non-anomalous $\mathbb{Z}_2$ in single character theories of different central charges. In \cite{Hohn:2023auw}, they classify self-dual orbifolds of single character CFTs with a central charge up to $24$. In this sense, they study the $\eta$ line. In this manuscript, we study the $\mathcal{N}$ line and provide defect partition functions and suitable crossing kernels by using the spin-selection rule. Our analysis is, therefore, similar to the analysis in \cite{Burbano:2021loy} but for the $c=24$ case. For ease of reading, when we discuss orbifolding with respect to the $\eta$ line, we have tried to make our notation compatible with \cite{BoyleSmith:2023xkd} who studied self-dual orbifolds for $c=16$ CFTs.

\section{Duality defects in $D_n$-type Niemeier lattice CFTs: key features}
\label{sec:D-def-key}
We will now highlight the key features in computing the duality defect insertion and defect Hilbert space partition functions for $D_n$-type Niemeier lattice CFTs by focusing on two particular cases. These cases will highlight the additional subtleties that arise compared to the $E_{8,1}$ CFT.

\subsection{Duality defects for $D_{24,1}$ }
We will first consider the case of $D_{24,1}$ CFT which has conjugacy classes $(0)_{D_{24}}, (s)_{D_{24}}$, and $N=1128$ in \eqref{j-function-character}. This lattice does not have outer automorphisms, as the outer automorphism of $(0)_{D_{24}}$ which switches the antenna of the $D_{24}$ Dynkin diagram, does not leave $(s)_{D_{24}}$ invariant. Therefore, we need to again look at inner automorphisms to find non-anomalous $\mathbb{Z}_2$ lines to orbifold with. As discussed earlier, we can use the Ka\v c-theorem to find such inner automorphisms. The affine $D_{24,1}$ Dynkin diagram has marks $a_i=2$ for $i=2,\ldots,22$, and $a_i=1$ for the rest. 
\begin{equation}
	\begin{tikzpicture}[baseline={(current bounding box.center)}, scale = 1]
		\tikzstyle{vertex}=[circle, fill=black, minimum size=2pt,inner sep=2pt];
		\def\r{1.2};
		\node[vertex] (T2) at (\r*3.134,\r*0.866) {};
		\node[vertex] (T3) at (\r*3.134,-1*\r*0.866) {};
		\node[vertex] (T4) at (\r*4,\r*0) {};
		\node[vertex] (T5) at (\r*5,\r*0) {};
		\node[vertex] (T6) at (\r*6,\r*0) {};
		\node[vertex] (T7) at (\r*7,\r*0) {};
		\node[vertex] (T8) at (\r*7.866,\r*0.866) {};
		\node[vertex] (T9) at (\r*7.866,-1*\r*0.866) {};
		
		\draw[-] (T2) -- (T4);
		\draw[-] (T3) -- (T4);
		\draw[-] (T4) -- (T5);
		\draw[-] (T7) -- (T6);
		\node[] (dots) at (\r*5.5,\r*0) {$\dots$};
		\draw[-] (T7) -- (T8);
		\draw[-] (T7) -- (T9);
		
		\draw[left] (T2) node {$\alpha_{0}$};
		\draw[left] (T3) node {$\alpha_1$};
		\draw[below] (T4) node {$\alpha_2$};
		\draw[below] (T5) node {$\alpha_3$};
		\draw[below] (T6) node {$\alpha_{21}$};
		\draw[below] (T7) node {$\alpha_{22}$};
		\draw[right] (T9) node {$\alpha_{24}$};
		\draw[right] (T8) node {$\alpha_{23}$};
	\end{tikzpicture}
\end{equation}
From the condition $\sum_i a_i s_i = 2$, we get the solutions: 
\begin{itemize}
	\item A single $s_r=1$ with $r \in \{1,2,\ldots,22\}$. In this case, $x=\frac{\omega_r}{2}$ where $\omega_r$ is the $r^{\text{th}}$ fundamental weight. This breaks $D_{24,1}$ into $D_{r,1} \times D_{24-r,1}$ after deleting the $r^{\text{th}}$ node in the affine Dynkin diagram. Since the inner automorphism is projectively realised, one may have a phase acting on the $(s)_{D_24}$ conjugacy class $\zeta_s=\pm 1$. We can define $\zeta_s=e^{\pi i \phi_s}$ with $\phi_s=0,1$. We will denote such inner automorphisms as $(r,\phi_s)$. Further, we can check that $x \cdot x \in \mathbb{Z}$ only if $r \in 4\mathbb{Z}$. Further, deleting the $r^{\text{th}}$ node and $(24-i)^{\text{th}}$ produces the same invariant subalgebra. Therefore, we only need to consider $r=4,8,12$. In these cases, one can also argue that $\phi_s=0$ and $\phi_s=1$ are conjugate to each other. Hence, we can focus on $\phi_s=0$ \cite{BoyleSmith:2023xkd}. Therefore, we have the cases $(4,0), (8,0), (12,0)$.
	\item Two of the $s_r$ are non-zero with $s_r,s_{r^\prime}$ where $i,j\in \{0,1,23,24\}$. We denote this as $(r,r^\prime,\zeta_s)$. In this case, $x=\frac{\omega_r+\omega_{r^\prime}}{2}$. Independent non-anamolous case corresponds to $(0,24,1)$  where $D_{24,1}\rightarrow A_{23,1}\times U(1)$. Note that $\omega_0=0$ in this formalism. Since this case contains $U(1)$, the lie algebra is not semi-simple, and this breaking is a reductive breaking as the rank reduces by $1$ for the semi-simple factors. We will exclude this case from our analysis. For $c=16$, this case has been studied in \cite{BoyleSmith:2023xkd}, and the analysis at $c=24$ would proceed similarly.
\end{itemize}
In this section, we will discuss the former cases to illustrate how to deal with the glue code while constructing the $\mathcal{N}$ defect line. Detailed results for all the cases are given in the next section.

Once we obtain an $x$-vector corresponding to a non-anamolous $\mathbb{Z}_2$, the orbifold partition function $\mathcal{Z}^{orb}$ is given as,
\begin{equation}\label{orbifold_partition_function}
	\mathcal{Z}^{\text{orb}}=\frac{1}{2}\left(\mathcal{Z}+\mathcal{Z}^{\eta}+\mathcal{Z}_{\eta}+\mathcal{Z}^{\eta}_{\eta}\right)\,.
\end{equation}
The partition function with the operator insertion in the partition function is,
\begin{equation}\label{partition_fn_with_spatial_ins}
	\mathcal{Z}^{\eta}=\frac{1}{(\eta(\tau))^{24}}\sum_{\alpha\in L} e^{2\pi ix\cdot\alpha}q^{\alpha\cdot\alpha/2}\,,
\end{equation}
where the translation of sigma model fields $X^i \rightarrow X^i+x^i$ has produced the phase $e^{2\pi i x\cdot \alpha}$ on states created by the vertex operators $V_\alpha$, and the $\frac{1}{(\eta(\tau))^{24}}$ is the oscillator contribution that is unchanged by this translation.

For the $(r,0)$ cases, the original lattice breaks into conjugacy classes of $D_r\times D_{24-r}$ as, 
\begin{equation}\label{420-decomp}
	L=(0)_{D_{24}}+(s)_{D_{24}}=(0_{D_{r}},0_{D_{24-r}})+(v_{D_{r}},v_{D_{24-r}})+(s_{D_{r}},s_{D_{24-r}})+(c_{D_{r}},c_{D_{24-r}}),
\end{equation}
as explained in appendix \ref{d-to-d-decomposition}. The vector $x=\frac{\omega_r}{2}$ is in the conjugacy class $(s_{D_{r}},0_{D_{24-r}})$. The phase factor for $i\in\{0,v,s,c\}$, can be calculated easily,

\begin{align}\label{d24-orb-phases}
	e^{2\pi i (x_s,\alpha)}=
	\begin{cases}
		1 &\text{ if } \alpha\in (0_{D_r},i_{D_{24-r}}) \, ,\\
		-1 	&\text{ if } \alpha\in (v_{D_r},i_{D_{24-r}})\, ,\\
		-1  &\text{ if } \alpha\in (c_{D_r},i_{D_{24-r}})\, ,\\
		1 &\text{ if } \alpha\in (s_{D_r},i_{D_{24-r}})\, .
	\end{cases}
\end{align}
It is immediately clear for $r=4, 8, 12$, that the invariant lattice $L_0$ under the inner automorphism symmetry is,
\begin{align}
L_0=(0_{D_{r}},0_{D_{24-r}})+(s_{D_{r}},s_{D_{24-r}}).
\end{align}
Similarly,
\begin{align}
L-L_0&=(v_{D_{r}},v_{D_{24-r}})+(c_{D_{r}},c_{D_{24-r}})\nonumber\\
(L-L_0)+x&=(c_{D_{r}},v_{D_{24-r}})+(v_{D_{r}},c_{D_{24-r}})\nonumber\\
L_0+x&=(s_{D_{r}},0_{D_{24-r}})+(0_{D_{r}},s_{D_{24-r}}).
\end{align}
For $r=12$, we find that $x\cdot x=3 \in 2\mathbb{Z}+1$, therefore $L^\prime=L_0 \cup \left((L-L_0)+x\right)$. Therefore the $D_{12,1}\times D_{12,1}$ breaking we have,
\begin{align}
L^\prime=(0_{D_{12}},0_{D_{12}})+(s_{D_{12}},s_{D_{12}})+(c_{D_{12}},v_{D_{12}})+(v_{D_{12}},c_{D_{12}}).
\end{align}
The above is nothing but the $D_{12,1}^2$ theory from the Schellekens list. In fact, one can see that the above glue code can be generated by using the generators given in Table \ref{Table-D-type-Schellekens} with $(s)_{D_{12}}\leftrightarrow (c)_{D_{12}}$ which is an equivalent description. Therefore, for the case $r=12$ the lattice description tells us that $\mathcal{Z}^{\text{orb}}=\mathcal{Z}^{D_{12,1}^2}=j-744+552$.  For the case $r=8$, we have $x\cdot x = 2 \in 2\mathbb{Z}$. Therefore $L^\prime = L_0\cup (L_0+x)$ which gives,
\begin{align}
	L^\prime&=(0_{D_{8}},0_{D_{16}})+(s_{D_{8}},s_{D_{16}})+(s_{D_{8}},0_{D_{16}})+(0_{D_{8}},s_{D_{16}})\nonumber\\
	&=(0_{D_{8}}+s_{D_{8}},0_{D_{16}})+(0_{D_{8}}+s_{D_{8}},s_{D_{16}})\nonumber\\
	&=(0_{E_{8}},0_{D_{16}})+(0_{E_{8}},s_{D_{16}}),
\end{align}
where we have used the fact that $0_{E_{8}}=0_{D_{8}}+s_{D_{8}}$. From this, we can see that the orbifolded theory is nothing but the $E_{8,1}D_{16,1}$ CFT from the Schellekens list. Therefore we expect $\mathcal{Z}^{\text{orb}}=\mathcal{Z}^{E_{8,1}D_{16,1}}=j-744+744$. For the case $r=4$, we have $x\cdot x=1 \in 2\mathbb{Z}+1$, therefore $L^\prime=L_0 \cup \left((L-L_0)+x\right)$ and this gives,
\begin{align}
L^\prime=(0_{D_{4}},0_{D_{20}})+(s_{D_{4}},s_{D_{20}})+(c_{D_{4}},v_{D_{20}})+(v_{D_{4}},c_{D_{20}}).
\end{align}
Note, however, that $D_4$ is special due to triality, and therefore, on the glue code, one can do the exchange $c_{D_{4}} \leftrightarrow v_{D_{4}}$ to achieve an equivalent glue code. This in fact takes us back to the original $D_{24,1}$ lattice in the $D_{4,1}D_{20,1}$ decomposition given in \eqref{420-decomp}. Therefore, we expect this case to correspond to a self-dual orbifold. We can verify these expectations from the lattice picture by explicitly calculating the orbifold partition function.

To obtain $\mathcal{Z}^{\text{orb}}$ we need to compute $\mathcal{Z}, \mathcal{Z}^\eta,\mathcal{Z}_\eta=S\mathcal{Z}^\eta$ and $\mathcal{Z}^\eta_\eta=T\mathcal{Z}_\eta$. The partition function without a defect insertion is,
\begin{equation}\label{D24_partition_fn}
	\mathcal{Z}=\frac{1}{2\eta^{24}(\tau)}\left(\theta_2^{24}+\theta_3^{24}+\theta_4^{24}\right)\, .
\end{equation}
To compute the partition function with the defect insertion we can use \eqref{partition_fn_with_spatial_ins} and \eqref{d24-orb-phases} to obtain,
\begin{align}\label{Z^eta_D24}
	\mathcal{Z}^{\eta}&=\frac{1}{\eta^{24}(\tau)}\left[\sum_{\alpha\in L_0}q^{(\alpha,\alpha)/2}+\sum_{\alpha \in L-L_0}e^{2\pi i (x,\alpha)}q^{(\alpha,\alpha)/2}\right]\, ,\nonumber\\
	&=\frac{1}{\eta^{24}(\tau)}\left[\sum_{\alpha\in (0_{D_r},0_{D_{24-r}})}q^{(\alpha,\alpha)/2}-\sum_{\alpha\in (v_{D_r},v_{D_{24-r}})}q^{(\alpha,\alpha)/2}\right]\, ,\nonumber\\
	&=\frac{1}{2\eta^{24}(\tau)}\left[\theta_3^r\theta_4^{24-r}+\theta_3^{24-r}\theta_4^{r}\right]\, .
\end{align}
The lattice vector sum is only over the lattice vectors $(0_{D_r},0_{D_{24-r}})$ and $(v_{D_r},v_{D_{24-r}})$ since the spinor and conjugate spinor cancel with each other according to \eqref{d24-orb-phases}. The modular-$S$ transformation $\tau\to -1/\tau$ of the $\theta(\tau)$ functions \eqref{s-t-theta} gives,
\begin{equation}\label{Z_eta_D24}
	\mathcal{Z}_{\eta}=\frac{1}{2\eta^{24}(\tau)}\left[\theta_3^r\theta_2^{24-r}+\theta_3^{24-r}\theta_2^{r}\right]\, .
\end{equation}
The modular-$T$ transformation $\tau\to\tau+1$,
\begin{equation}
	\mathcal{Z}_{\eta}^{\eta}=\frac{1}{2\eta^{24}(\tau)}\left[e^{-i r \pi/4}\theta_4^r\theta_2^{24-r}+e^{i r \pi/4}\theta_4^{24-r}\theta_2^{r}\right]\, .
\end{equation}
Since $r=4,8,12$ for us, $e^{-i r \pi/4}=e^{i r \pi/4}$. Therefore,
\begin{equation}\label{Z^eta_eta_D24}
	\mathcal{Z}_{\eta}^{\eta}=\frac{1}{2\eta^{24}(\tau)}e^{i r \pi/4}\left[\theta_4^r\theta_2^{24-r}+\theta_4^{24-r}\theta_2^{r}\right]\,,
\end{equation}
with $e^{i r \pi/4}=1$ for $r=8$ and $e^{i r \pi/4}=-1$ for $r=4,12$. The partition function can be evaluated as $q$-series expansion whose level 1 coefficient gives the number of generators as usual.  The partition function of the orbifold theory for these anomaly free cases corresponds to the following current algebra CFTs,
\begin{align}		
r=4&\implies D_{24,1} \, ,\nonumber\\
r=8&\implies D_{16,1}E_{8,1}\, ,\nonumber\\
r=12&\implies (D_{12,1})^2 \,,
\end{align}\label{D24-duality-web}
as expected from the lattice description. Thus, $r=4$ corresponds to a self-dual $\mathbb{Z}_2$ orbifold of $D_{24,1}$ CFT, and we can investigate for the duality defect line $\mathcal{N}$. Before doing so, let us consider the orbifolding that Ka\v c-theorem does not give. These automorphisms are the case where we do not delete any Dynkin nodes, but we give a projective phase $-1$ to the vertex operators in the conjugacy class $(s)_{D_{24}}$. This case is denoted as $(0,\phi_s=1)$. The $\eta$ insertion partition function can be calculated by simply having a negative sign in front of the contribution by the $(s)_{D_{24}}$ conjugacy class, and we get,
\begin{equation}
	\mathcal{Z}^\eta= \frac{1}{2\eta^{24}}\left[\theta_3^{24}+\theta_4^{24}-\theta_2^{24}\right]\, ,
\end{equation}
and the modular transformation gives,
\begin{equation}
	\mathcal{Z}_{\eta}= \frac{1}{2\eta^{24}}\left[\theta_3^{24}+\theta_2^{24}-\theta_4^{24}\right]\, .
\end{equation}
The modular-$T$ transformation of the above is unambiguous and leads to,
\begin{equation}
	\mathcal{Z}^\eta_{\eta}= \frac{1}{2\eta^{24}}\left[\theta_4^{24}+\theta_2^{24}-\theta_3^{24}\right]\, .
\end{equation}
From the above we can calculate the orbifold partition function to be,
\begin{equation}
	\mathcal{Z}^{orb}=\frac{1}{2\eta^{24}}\left[\theta_2^{24}+\theta_3^{24}+\theta_4^{24}\right]\, .
\end{equation}
Thus, we have another self-dual orbifold generated by the symmetry $(0,1)$. This is in fact analogous to the self-dual $\mathbb{Z}_2$ orbifold for the $E_{8,1} \rightarrow D_{8,1}$ case considered in \cite{Burbano:2021loy} if we began from the outset by considering the $(0)_{E_8}$ lattice as $(0)_{D_8}+(s)_{D_8}$. Then the automorphism that would leave the invariant lattice $L_0=(0)_{D_8}$ would be $(0,1)$ in the $D_8$ description. We will now proceed to construct the duality defect for both cases $(4,0)$ and $(0,1)$ in $D_{24,1}$ CFT. 

\paragraph{Duality defect in the $(4,0)$ case:}

We consider the self-dual orbifold $D_4\times D_{20} \subset D_{24}$ and construct the duality defects explicitly. As we have already discussed, in this case,
\begin{align}
L_0=(0_{D_{4}},0_{D_{20}})+(s_{D_{4}},s_{D_{20}}).
\end{align}
We consider the outer automorphism of the above lattice that implements $(L-L_0)\leftrightarrow  \left((L-L_0)+x\right)$. The only Dynkin diagram symmetry which leaves the $L_0$ sector invariant is the $\mathbb{Z}_2$ transformation $(v)_{D_4}\leftrightarrow (c)_{D_4}$ as shown in the Dynkin diagram below. The roots transform as $\alpha_1 \leftrightarrow \alpha_3$
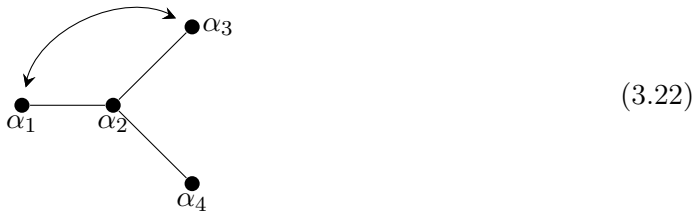
\begin{equation}
	\begin{tikzpicture}[baseline={(current bounding box.center)}, scale = 1]
		\tikzstyle{vertex}=[circle, fill=black, minimum size=2pt,inner sep=2pt];
		\def\r{1.2};
		\node[vertex] (T1) at (\r*5,\r*0) {};
		\node[vertex] (T2) at (\r*6,\r*0) {};
		\node[vertex] (T3) at (\r*6.866,\r*0.866) {};
		\node[vertex] (T4) at (\r*6.866,-1*\r*0.866) {};
		
		\draw[-] (T1) -- (T2);
		\draw[-] (T2) -- (T3);
		\draw[-] (T2) -- (T4);
		\draw[below] (T1) node {$\alpha_1$};
		\draw[below] (T2) node {$\alpha_2$};
		\draw[right] (T3) node {$\alpha_3$};
		\draw[below] (T4) node {$\alpha_4$};
		\draw [->- = 0.95 rotate 0, ->- = 0.05 rotate 180, black, shorten >=5pt, shorten <=5pt] (T1) to [out=80,in=150] (T3);
	\end{tikzpicture}
\label{d4v-d4c}
\end{equation}
Thus the invariant lattice is composed of the root lattice corresponding to $C_3$ algebra,
\begin{equation}
	\begin{tikzpicture}[baseline={(current bounding box.center)}, scale = 1]
		\tikzstyle{vertex}=[circle, fill=black, minimum size=2pt,inner sep=2pt];
		\def\r{1.2};
		\node[vertex] (T1) at (\r*5,\r*0) {};
		\node[vertex] (T2) at (\r*6,\r*0) {};
		\node[vertex] (T3) at (\r*7.5,\r*0) {};
		
		\draw[-] (T1) -- (T2);
		\draw[below] (T1) node {$\alpha_4^{(D_4)}$};
		\draw[below] (T2) node {$\alpha_2^{(D_4)}$};
		\draw[below] (T3) node {$(\alpha_1^{(D_4)}+\alpha_3^{(D_4)})$};
		\draw[above] (T1) node {$\alpha_1^{(C_3)}$};
		\draw[above] (T2) node {$\alpha_2^{(C_3)}$};
		\draw[above] (T3) node {$\alpha_3^{(C_3)}$};
		\draw[-r- = 0.60 rotate 0, double,double distance = 0.05cm,double,double distance = 0.05cm] (T3) -- (T2);
	\end{tikzpicture}
\end{equation}\label{C3-Dynkin}
To obtain the $C_3$ description of the $(s)_{D_4}$ conjugacy class that is a part of $L_0$ we need to write the fundamental weight corresponding to $(s)_{D_4}$,
\begin{align}
	\omega_{4}^{(D_4)}=\frac{1}{2}(\alpha_1^{(D_4)}+\alpha_3^{(D_4)})+\alpha_2^{(D_4)}+\alpha_4^{(D_4)}\in(s)_{D_4}\, .
\end{align}
This remains invariant under the outer automorphism switch $\alpha_1 \leftrightarrow \alpha_3$ as required. In the above equation, the last two terms correspond to the simple roots of $C_3$ as can be clearly seen from \eqref{C3-Dynkin}. and therefore they belong to the root conjugacy class. In fact, in terms of the fundamental weight of the $C_3$ lattice,
\begin{equation}
	\omega_{4}^{(D_4)}=\omega_{1}^{(C_3)}\, .
\end{equation}
This fixes the invariant lattice to be the $(0_{C_3},0_{D_{20}})+(1_{C_3},s_{D_{20}})$, where $1_{C_3}$ is the conjguacy class corresponding to $\omega_{1}^{(C_3)}$. Knowing the invariant lattice fixes the oscillator contributions to be $\frac{1}{\eta_g(\tau)}$ where $\eta_g(\tau)=\eta(\tau)^{22}\eta(2\tau)$. One could, however, have additional phases to the vertex operators, which will contribute to the lattice $\theta$ functions. These phases are $e^{2\pi i x\cdot \alpha}$ for vertex operator $V_{\alpha}$, where $x$ is given by Ka\v c theorem for outer automorphisms, so one has to consider twisted Dynkin diagrams with $k=2,3$. For $\mathbb{Z}_2$ outer automorphisms, we can look at $k=2$. Keeping aside the $D_{20}$ sector, the automorphism vector $x$ is given in terms of the fundamental weights of $B_3$ lattice, which is dual of the invariant lattice $C_3$. The two possible choices for the automorphism vector are given by Ka\v c theorem to be,
\begin{itemize}
	\item $x_1=0$, invariant subalgebra is $B_3$\, ,\hspace{2.8 cm}\begin{tikzpicture}[baseline={(current bounding box.center)}, scale = 1]
		\tikzstyle{vertex}=[circle, fill=black, minimum size=2pt,inner sep=2pt];
		\def\r{1.2};
		\node[cross out,draw] (T1) at (\r*5,\r*0) {};
		\node[vertex] (T2) at (\r*6,\r*0) {};
		\node[vertex] (T3) at (\r*7,\r*0) {};
		\node[vertex] (T4) at (\r*8,\r*0) {};
		
		\draw[-, dashed] (T1) -- (T2);
		\draw[-] (T2)--(T3);
		\draw[-r- = 0.40 rotate 180, double,double distance = 0.05cm,double,double distance = 0.05cm] (T4) -- (T3);
	\end{tikzpicture}
	\item $x_2=\frac{1}{2} \omega_1^{(B_3)}$ with invariant subalgebra $A_1\oplus B_2$\, \qquad \begin{tikzpicture}[baseline={(current bounding box.center)}, scale = 1]
		\tikzstyle{vertex}=[circle, fill=black, minimum size=2pt,inner sep=2pt];
		\def\r{1.2};
		\node[vertex] (T1) at (\r*5,\r*0) {};
		\node[cross out,draw] (T2) at (\r*6,\r*0) {};
		\node[vertex] (T3) at (\r*7,\r*0) {};
		\node[vertex] (T4) at (\r*8,\r*0) {};
		
		\draw[-, dashed] (T1) -- (T2);
		\draw[-,dashed] (T2)--(T3);
		\draw[-r- = 0.40 rotate 180, double,double distance = 0.05cm,double,double distance = 0.05cm] (T4) -- (T3);
	\end{tikzpicture}
\end{itemize}
In the above we have indicated the invariant subalgebra after the phases are inserted. 

\paragraph{Computing duality defect insertion and defect Hilbert space partition functions:}
We will now calculate the defect insertion partition function $\mathcal{Z}^\mathcal{N}$ and the defect Hilbert space partition function $\mathcal{Z}_\mathcal{N}$ for both the $x_1,x_2$ cases above.

\vspace{3mm}
\noindent\colorbox{silver}{$x_1=0$}:\\ 
The insertion partition function is given as,
\begin{equation}\label{C3-before-D3}
	\mathcal{Z}[D_{24}]^{\mathcal{N}}_1=\frac{\sqrt{2}}{\eta_g(\tau)}\left[\chi_0^{D_{20}}\sum_{\alpha\in (0)_{C_3}}q^{(\alpha,\alpha)/2}+\chi_s^{D_{20}}\sum_{\beta\in (1)_{C_3}}q^{(\beta,\beta)/2}\right]\, ,
\end{equation}
where the conjugacy class $(1)_{C_3}$ is defined by the relation $\omega_{1}^{(C_3)}=(1,0,0)\in (1)_{C_3}$ and $\sqrt{2}$ is added as usual to get the right fusion rules.  We can use the equivalence between the root lattices $\Lambda_{C_3}\equiv\Lambda_{D_3}$ to compute the above partition function. Note that the root lattice in $D_3$ is explicitly given as $\alpha=(k_1,k_2,k_3)\in \Lambda_{D_3}$ with $k_1,k_2,k_3 \in\mathbb{Z}$ such that $\sum_{k_i\in\mathbb{Z}}k_i=0$ mod $2$, $i=1,2,3$. The conjugacy class $(1)_{C_3}$ is obtained as $\Lambda_{D_3}+(1,0,0)$. The sum in \eqref{C3-before-D3} can now be written in terms of sum over $D_3$ vectors,
\begin{align}
	\mathcal{Z}[D_{24}]^{\mathcal{N}}_1&=\left(\frac{\theta_3^{20}(\tau)+\theta_4^{20}(\tau)}{2\eta^{20}(\tau)}\right)\frac{\sqrt{2}}{\eta_g(\tau)}\left[\sum_{k_i\in\mathbb{Z}} \left(\frac{1+(-1)^{\sum k_i}}{2}\right)q^{\frac{k_1^2+k_2^2+k_3^2}{2}}\right]\nonumber\\
	&\quad+\left(\frac{\theta_2^{20}(\tau)}{2\eta^{20}(\tau)}\right)\frac{\sqrt{2}}{\eta_g(\tau)}\left[\sum_{k_i\in\mathbb{Z}} \left(\frac{1+(-1)^{\sum k_i}}{2}\right)q^{\frac{(k_1+1)^2+k_2^2+k_3^2}{2}}\right]\, ,\nonumber\\
	&=\sqrt{2}\left(\frac{\theta_3^{20}(\tau)+\theta_4^{20}(\tau)}{4\eta^{24}(\tau)}\right)\left[\theta_3^{3+1/2}(\tau)\theta_4^{1/2}(\tau)+\theta_4^{3+1/2}(\tau)\theta_3^{1/2}(\tau)\right]\nonumber\\
	&\quad+\sqrt{2}\left(\frac{\theta_2^{20}(\tau)}{4\eta^{24}(\tau)}\right)\left[\theta_3^{3+1/2}(\tau)\theta_4^{1/2}(\tau)-\theta_4^{3+1/2}(\tau)\theta_3^{1/2}(\tau)\right]\, .
\end{align}
This is modular $T$-transformation invariant. In terms of the characters of $so(n)$,
\begin{equation}\label{Z1-D24-insertion}
	\mathcal{Z}[D_{24}]^{\mathcal{N}}_1=\sqrt{2}\chi_{0}^{so(40)}\left(\chi_{0}^{so(7)}\chi_0^{so(1)}-\chi_{v}^{so(7)}\chi_v^{so(1)}\right)+\sqrt{2}\chi_{s}^{so(40)}\left(\chi_{v}^{so(7)}\chi_0^{so(1)}-\chi_{0}^{so(7)}\chi_v^{so(1)}\right)\, .
\end{equation}
The modular $S$-transformation of the above partition function should have positive coefficients when expanded in terms of the characters since it is a partition function on a twisted Hilbert space.
\begin{equation}
	\mathcal{Z}[D_{24}]_ {\mathcal{N},\;  1}=\sqrt{2}\left(\chi_0^{so(7)}+\chi_v^{so(7)}\right)\chi_s^{so(1)}\chi_0^{so(40)}+\sqrt{2}\left(\chi_0^{so(1)}+\chi_v^{so(1)}\right)\chi_s^{so(7)}\chi_v^{so(40)}\, .
\end{equation}
The $q$-expansion indeed gives positive integer degeneracies in this case. Further, the conformal dimensions in the defect Hilbert space satisfy the spin selection rule appropriate for the duality defect Hilbert space with $h=\pm \frac{1}{16}+\frac{\mathbb{Z}}{2}$ and therefore $\epsilon=1$ in the fusion kernel and the duality defect line $\mathcal{N}$ is in the same category as that of the duality defect in Ising CFT.

\vspace{3mm}
\noindent\colorbox{silver}{$x_2=\frac{1}{2}\omega_1^{B_3}$}:\\
We can now discuss the case with $x_2=\frac{1}{2}\omega_1^{B_3}$.  The defect insertion partition function is given by\footnote{The sub-script $3$ on the partition function is written for later convenience where we write the expression for the defect insertion partition function for the $x_1,x_2$ case together succinctly.},
\begin{equation}
	\mathcal{Z}[D_{24}]^{\mathcal{N}}_3=\frac{\sqrt{2}}{\eta_g(\tau)}\left[\chi_0^{D_{20}}\sum_{\alpha\in (0)_{C_3}}e^{2\pi i (x_2,\alpha)}q^{(\alpha,\alpha)/2}+\chi_s^{D_{20}}\sum_{\beta\in (1)_{C_3}}e^{2\pi i (x_2,\beta)}q^{(\beta,\beta)/2}\right]\, .
\end{equation}
To compute, we need to evaluate the two inner products in the above partition function.  We can write the inner product of $x_2$ with the general root vector in $\Lambda_{C_3}$ as,
\begin{align}
(\omega_{1}^{B_3},\alpha_{C_3})&=k_1\in\mathbb{Z}\,,
\end{align}
and with the $(1)_{C_3}$ conjugacy class as,
\begin{align}
(\omega_{1}^{B_3},\omega_1^{C_3})&=1.
\end{align}
This implies ,
\begin{align}
	e^{2\pi i(x_2,\alpha)}&=e^{\pi i(\omega_1^{B_3},\alpha)}=(-1)^{k_1}, \quad \alpha\in (0)_{C_3}\, ,\nonumber\\
	e^{2\pi i(x_2,\beta)}&=e^{\pi i(\omega_1^{B_3},\alpha)}e^{\pi i(\omega_1^{B_3},\omega_{1}^{C_3})}=-(-1)^{k_1}, \quad \alpha\in (0)_{C_3}\, .
\end{align}
We can subtitute this back in the insertion partition function and again use the $D_3$ lattice descripton to write, 
\begin{align}
	\mathcal{Z}[D_{24}]^{\mathcal{N}}_3&=\left(\frac{\theta_3^{20}(\tau)+\theta_4^{20}(\tau)}{2\eta^{20}(\tau)}\right)\frac{\sqrt{2}}{\eta_g(\tau)}\left[\sum_{k_i\in\mathbb{Z}} \left(\frac{1+(-1)^{\sum k_i}}{2}\right)(-1)^{k_1}q^{k_1^2/2+k_2^2/2+k_3^2/2}\right]\nonumber\\
	&\quad-\left(\frac{\theta_2^{20}(\tau)}{2\eta^{20}(\tau)}\right)\frac{\sqrt{2}}{\eta_g(\tau)}\left[\sum_{k_i\in\mathbb{Z}} \left(\frac{1+(-1)^{\sum k_i}}{2}\right)(-1)^{k_1}q^{(k_1+1)^2/2+k_2^2/2+k_3^2/2}\right]\, ,\nonumber\\
   &=\sqrt{2}\left(\frac{\theta_3^{20}(\tau)+\theta_4^{20}(\tau)}{4\eta^{24}(\tau)}\right)\left[\theta_3^{2+1/2}(\tau)\theta_4^{1+1/2}(\tau)+\theta_4^{2+1/2}(\tau)\theta_3^{1+1/2}(\tau)\right]\nonumber\\
	&\quad +\sqrt{2}\left(\frac{\theta_2^{20}(\tau)}{4\eta^{24}(\tau)}\right)\left[\theta_3^{2+1/2}(\tau)\theta_4^{1+1/2}(\tau)-\theta_4^{2+1/2}(\tau)\theta_3^{1+1/2}(\tau)\right]\, .
\end{align}
In terms of the characters of $so(n)$,
\begin{equation}
	\mathcal{Z}[D_{24}]^{\mathcal{N}}_3=\sqrt{2}\left(\chi_{0}^{so(40)}\left(\chi_{0}^{so(5)}\chi_0^{so(3)}-\chi_{v}^{so(5)}\chi_v^{so(3)}\right)+\chi_{s}^{so(40)}\left(\chi_{v}^{so(5)}\chi_0^{so(3)}-\chi_{0}^{so(5)}\chi_v^{so(3)}\,\right) \right),
\end{equation}
and under modular $S$-transformation,
\begin{equation}
	\mathcal{Z}[D_{24}]_{ \mathcal{N},\; 3}=\sqrt{2}\left(\left(\chi_0^{so(5)}+\chi_v^{so(5)}\right)\chi_s^{so(3)}\chi_0^{so(40)}+\left(\chi_0^{so(3)}+\chi_v^{so(3)}\right)\chi_s^{so(5)}\chi_v^{so(40)}\right)\, .
\end{equation}
This final partition function is over the defect Hilbert space, and again, one can check that the $q$ expansion has positive integer coefficients, which correspond to degeneracies. Futher, the conformal dimensions $h \in \pm \frac{3}{16}+\frac{\mathbb{Z}}{2}$, and therefore the duality defect has the fusion kernel with $\epsilon=-1$ and is in the same category as $\text{Ising}\times SU(2)_1 \text{WZW model}$.

\vspace{3mm}
\noindent\colorbox{silver}{Final result}:
The two defect partition functions can be written succinctly as,
\begin{empheq}[box=\widefbox]{align}   \mathcal{Z}[D_{24}]^{\mathcal{N}}_i&=\sqrt{2}\left(\chi_{0}^{so(40)}\left(\chi_{0}^{so(8-i)}\chi_0^{so(i)}-\chi_{v}^{so(8-i)}\chi_v^{so(i)}\right)\right.\nonumber\\ &\left.  +\chi_{s}^{so(40)}\left(\chi_{v}^{so(8-i)}\chi_0^{so(i)}-\chi_{0}^{so(8-i)}\chi_v^{so(i)}\right)\right)\, ,
\end{empheq}

where $i=1,3$ for $x_1$ and $x_2$ respectively. 

\paragraph{Duality defect in the $(0,1)$ case:}

We now move to compute the defect partition function for the $(0,1)$ case. The outer automorphism which leaves the untwisted lattice $(0)_{D_{24}}$ invariant is the only Dynkin diagram symmetry of $D_{24}$, which is $(s)_{D_{24}}\leftrightarrow (c)_{D_{24}}$.
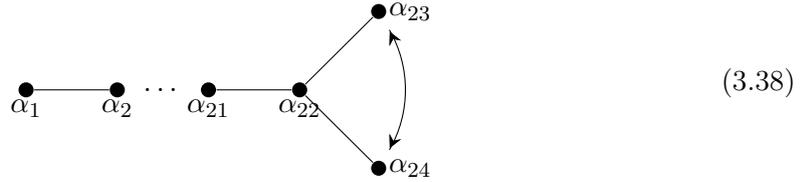
\begin{equation}
	\begin{tikzpicture}[baseline={(current bounding box.center)}, scale = 1]
		\tikzstyle{vertex}=[circle, fill=black, minimum size=2pt,inner sep=2pt];
		\def\r{1.2};
		\node[vertex] (T3) at (\r*3,\r*0) {};
		\node[vertex] (T4) at (\r*4,\r*0) {};
		\node[vertex] (T5) at (\r*5,\r*0) {};
		\node[vertex] (T6) at (\r*6,\r*0) {};
		\node[vertex] (T7) at (\r*6.866,\r*0.866) {};
		\node[vertex] (T8) at (\r*6.866,-1*\r*0.866) {};
		
		\draw[-] (T3) -- (T4);
		\node[] (dots) at (\r*4.5,\r*0) {$\dots$};
		\draw[-] (T5) -- (T6);
		\draw[-] (T6) -- (T7);
		\draw[-] (T6) -- (T8);
		
		\draw [->- = 0.9 rotate 0, ->- = 0.1 rotate 180, black, shorten >=5pt, shorten <=5pt] (T7) to [out=-60,in=60] (T8);
		\draw[below] (T3) node {$\alpha_1$};
		\draw[below] (T4) node {$\alpha_2$};
		\draw[below] (T5) node {$\alpha_{21}$};
		\draw[below] (T6) node {$\alpha_{22}$};
		\draw[right] (T7) node {$\alpha_{23}$};
		\draw[right] (T8) node {$\alpha_{24}$};
	\end{tikzpicture}
\end{equation}
The defect partition function is computed similar to the calculation above and is obtained to be,
\begin{equation}
	\boxed{	\mathcal{Z}[D_{24}]^{\mathcal{N}}_i=\sqrt{2}\left(\chi_{0}^{so(48-i)}\chi_0^{so(i)}-\chi_{v}^{so(48-i)}\chi_v^{so(i)}\right)}\, ,
\end{equation}
where, $i=1,3,5,\cdots,23$ obtained by cancelling nodes in the twisted $k=2$ Dynkin diagram of $D_{24}$. The form of the defect partition function is similar to the $\mathbb{Z}_2$ defect partition function in $E_8$ CFT \cite{Burbano:2021loy}. 

\subsection{Duality defects for $D_{12,1}^2$}

The glue code for the $D_{12}^2$ lattice CFT is,
\begin{equation}
(0_{D_{12}},0_{D_{12}})+(c_{D_{12}},c_{D_{12}})+(s_{D_{12}},v_{D_{12}})+(v_{D_{12}},s_{D_{12}})\,. 
\end{equation}
The independent non-anomalous automorphisms of the $D_{12}^2$ CFT and the corresponding orbifolded theories are given in the table below.
\begin{table}
\begin{center}
	\begin{tabular}{|c|c|}
	\hline
	Automorphism & Algebra for $\mathcal{Z}^{\text{orb}}$\\
	\hline
\rowcolor{lightcyan}
	$(0,0), (0,1)$ & $D_{12,1}^2$ \\
	\hline
	\rowcolor{lightcyan}
	$(0,0),(4,0)$ &  $D_{12,1}^2$\\
	\hline
	$(0,1),(0,1)$ &  $D_{24,1}$\\
	\hline
	$(0,1),(4,0)$& $D_{16,1}E_{8,1}$\\
	\hline
	\rowcolor{lightcyan}
	$(2,0),(2,0)$ &  $D_{12,1}^2$\\
	\hline
	$(2,0),(6,0)$ & $D_{10,1}E_{7,1}^2$ \\
	\hline
	$(4,0),(4,0)$& $D_{8,1}^3$ \\
	\hline
	$(6,0),(6,0)$ & $D_{6,1}^4$ \\
	\hline
\end{tabular}
\caption{\label{orbifolds_d12^2} The non-anomalous orbifolds of the $D_{12}^2$ Niemeier lattice CFT obtained from Ka\v c's theorem. The self-dual orbifolds are marked in blue.}
\end{center}
\end{table}
In the above, we have identified the orbifolded theory with the help of the $q$-expansion, as well as the lattice description, which helps break the degeneracy when there are multiple possible algebras for the same value of $N$ in the Schellekens list.

As we see above, the theory has three independent non-anomalous self-dual orbifolds with semi-simple factors. To denote the automorphism with respect to which we perform the orbifolding, we will use the same notation as earlier, but now given for both the $D_{12,1}$ factors. For instance, for single node deletion, we will have the notation $(r,\phi_s),(r^\prime,\phi^\prime_s)$ with $\phi_s=0$ if $r \; \text{or}\; r^\prime \neq 0$. The non-anomalous self-dual orbifolds for the non-reductive cases are,
\begin{itemize}
	\item $(0,0),(0,1)$. This corresponds to the case where both the $D_{12,1}$ have a trivial $x$-vector but for one of the $D_{12,1}$, the (conjugate-)spinor conjugacy class vertex operators acquire a projective phase.
	\item $(0,0),(4,0)$, with the breaking $D_{12}D_{12}\rightarrow D_{12}D_{8}D_{4}$. Here too, similar to the $D_{24}\rightarrow D_4D_{20}$ case, the triality of $D_4$ allows one to have a self-dual orbifold. 
	\item $(2,0),(2,0)$, with the breaking $D_{12}D_{12}\rightarrow D_{10}^{(1)}D_{2}^{(1)}D_{10}^{(2)}D_{2}^{(2)}$, where we have given superscripts to distinguish between the different $D_2$ and $D_{10}$ factors.
\end{itemize}
Among the above, the $(0,0),(0,1)$ case is similar to the $(0,1)$ computation of the $D_{24,1}$ CFT performed in the previous sub-section, as well as the $\mathbb{Z}_2$ self-dual orbifold of $E_{8,1}$ performed in \cite{Burbano:2021loy}. The second case $(0,0),(4,0)$ is similar to the $(4,0)$ case in $D_{24,1}$ performed in the previous sub-section. We will therefore focus on the $(2,0),(2,0)$ case and construct the duality defect there.

The $(2,0),(2,0)$ symmetry of the CFT is a special $\mathbb{Z}_2$ symmetry since it belongs to the $\mathbb{Z}_2$ subgroup of the $\mathbb{Z}_2\times \mathbb{Z}_2$ symmetries of the $D_{12}^2$ current algebra CFT, i.e, both $(2,0)$s acting on the two factors $D_{12}^{(1)}$ and $D_{12}^{(2)}$. The invariant lattice is,
\begin{align}
L_0=&(0_{D_{10}},0_{D_2},0_{D_{10}},0_{D_2})\oplus(v_{D_{10}},v_{D_2},v_{D_{10}},v_{D_2})\oplus(s_{D_{10}},s_{D_2},0_{D_{10}},v_{D_2})\nonumber\\
&\oplus(c_{D_{10}},c_{D_2},v_{D_{10}},0_{D_2})\oplus(v_{D_{10}},0_{D_2},c_{D_{10}},c_{D_2})\oplus(0_{D_{10}},v_{D_2},s_{D_{10}},s_{D_2})\nonumber\\
&\oplus(c_{D_{10}},s_{D_2},c_{D_{10}},s_{D_2})\oplus(s_{D_{10}},c_{D_2},s_{D_{10}},c_{D_2})\,.
\end{align}
The automorphism vector $x\in (0_{D_{10}},s_{D_2},0_{D_{10}},s_{D_2})$ and has odd norm. Therefore, we will have $L^\prime=L_0 \cup ((L-L_0)+x)$. However, when we try to find an outer automorphism that takes us from $L$ to $L^\prime$ we do not find any in the $D_{10}D_2D_{10}D_2$ description where the $D_2$ outer automorphism is taken to be $(s)_{D_2}\leftrightarrow(c)_{D_2}$ alone analogous to the higher $D_r$ case with $r\neq 4$. However we can consider the $A_1 \oplus A_1$ description of the $D_2$ algebra and write $(s)_{D_{2}}=(0_{A_1},1_{A_1})$ and $(c)_{D_{2}}=(1_{A_1},0_{A_1})$, where $1_{A_1}$ is the fundamental representation of $A_1$. Then $(v)_{D_{2}}=(1_{A_1},1_{A_1})$. The outer automorphism that changes $L$ to $L^\prime$ is exchange of two $A_1$s in the $D_{10}^{(1)}A_{1}^{(1)}A_{1}^{(2)}D_{10}^{(2)}A_{1}^{(3)}A_{1}^{(4)}$ description. We can consider, say, $A_{1}^{(2)}\leftrightarrow A_{1}^{(4)}$. The defect insertion partition function can be evaluated by noting that only states that are symmetric with respect to $A_{1}^{(2)},A_{1}^{(4)}$ contribute, which leads to a lattice theta function with $\tau \rightarrow 2\tau$. The defect insertion partition function for this exchange defect is,
\begin{align}
	\hspace{-4mm}\mathcal{Z}[D_{12}^2]^{\, \mathcal{N}}&=\chi_0^{D_{10}}(\tau)\chi_0^{A_1}(\tau)\chi_0^{A_1}(2\tau)\chi_0^{D_{10}}(\tau)\chi_0^{A_1}(\tau)+\chi_v^{D_{10}}(\tau)\chi_1^{A_1}(\tau)\chi_1^{A_1}(2\tau)\chi_v^{D_{10}}(\tau)\chi_1^{A_1}(\tau)\nonumber\\
	&+\chi_s^{D_{10}}(\tau)\chi_0^{A_1}(\tau)\chi_1^{A_1}(2\tau)\chi_0^{D_{10}}(\tau)\chi_1^{A_1}(\tau)+\chi_c^{D_{10}}(\tau)\chi_1^{A_1}(\tau)\chi_0^{A_1}(2\tau)\chi_v^{D_{10}}(\tau)\chi_0^{A_1}(\tau)\nonumber\\
	&+\chi_v^{D_{10}}(\tau)\chi_0^{A_1}(\tau)\chi_0^{A_1}(2\tau)\chi_c^{D_{10}}(\tau)\chi_1^{A_1}(\tau)+\chi_0^{D_{10}}(\tau)\chi_1^{A_1}(\tau)\chi_1^{A_1}(2\tau)\chi_s^{D_{10}}(\tau)\chi_0^{A_1}(\tau)\nonumber\\
	&+\chi_c^{D_{10}}(\tau)\chi_0^{A_1}(\tau)\chi_1^{A_1}(2\tau)\chi_c^{D_{10}}(\tau)\chi_0^{A_1}(\tau)+\chi_s^{D_{10}}(\tau)\chi_1^{A_1}(\tau)\chi_0^{A_1}(2\tau)\chi_s^{D_{10}}(\tau)\chi_1^{A_1}(\tau)\, .
\end{align}
The characters of $A_1$ can be written in terms of the usual Jacobi theta functions by utilising the relations with the characters of $\mathfrak{so}(4)$	,
\begin{align}
	\chi_0^{A_1}(\tau)=\frac{\left(\theta_3^2(\tau)+\theta_4^2(\tau)\right)^{1/2}}{\sqrt 2 \eta(\tau)}\, ,\nonumber\\
	\chi_1^{A_1}(\tau)=\frac{\left(\theta_3^2(\tau)-\theta_4^2(\tau)\right)^{1/2}}{\sqrt 2 \eta(\tau)}\, .
\end{align}
To obtain the modular-$S$ transformation of this partition function we use the doubling identities of the Jacobi theta functions as noted below,
\begin{align}
	\theta_2^2(2\tau)&=\frac{\theta_3^2(\tau)-\theta_4^2(\tau)}{2}\, ,\nonumber\\
	\theta_3^2(2\tau)&=\frac{\theta_3^2(\tau)+\theta_4^2(\tau)}{2}\, ,\nonumber\\
	\theta_4^2(2\tau)&=\theta_3(\tau)\theta_4(\tau)\, ,\nonumber\\
	\eta(2\tau)&=\frac{\eta ^2(\tau)}{\sqrt{\theta _3(\tau) \theta _4(\tau)}}\, .
\end{align}
The defect Hilbert space has the partition function in terms of the theta functions,
\begin{align}
  \mathcal{Z}[D_{12}^2]_{\mathcal{N}}&=\frac{\sqrt{\theta _2 \theta _3} \sqrt{\theta _2 \theta _3+\frac{1}{2} \left(\theta _2^2+\theta _3^2\right)} \left(\theta _3^2-\theta _2^2\right) \theta _4^{20}}{8 \sqrt{2} \eta ^{24}}\nonumber\\
  &+\frac{\sqrt{\theta _2 \theta _3} \sqrt{\frac{1}{2} \left(\theta _2^2+\theta _3^2\right)-\theta _2 \theta _3} \left(\theta _2^2+\theta _3^2\right) \theta _4^{20}}{8 \sqrt{2} \eta ^{24}}\nonumber\\
	&+\frac{\sqrt{\theta _2 \theta _3} \sqrt{\theta _3^2-\theta _2^2} \sqrt{\theta _2^2+\theta _3^2} \sqrt{\theta _2 \theta _3+\frac{1}{2} \left(\theta _2^2+\theta _3^2\right)} \left(\theta _3^{10}-\theta _2^{10}\right) \theta _4^{10}}{4 \sqrt{2} \eta ^{24}}\nonumber\\
	&+\frac{\sqrt{\theta _2 \theta _3} \sqrt{\theta _3^2-\theta _2^2} \sqrt{\theta _2^2+\theta _3^2} \sqrt{\frac{1}{2} \left(\theta _2^2+\theta _3^2\right)-\theta _2 \theta _3} \left(\theta _2^{10}+\theta _3^{10}\right) \theta _4^{10}}{4 \sqrt{2} \eta ^{24}}\nonumber\\
	&+\frac{\sqrt{\theta _2 \theta _3} \sqrt{\frac{1}{2} \left(\theta _2^2+\theta _3^2\right)-\theta _2 \theta _3} \left(\theta _3^2-\theta _2^2\right) \left(\theta _3^{10}-\theta _2^{10}\right){}^2}{8 \sqrt{2} \eta ^{24}}\nonumber\\
	&+\frac{\sqrt{\theta _2 \theta _3} \sqrt{\theta _2 \theta _3+\frac{1}{2} \left(\theta _2^2+\theta _3^2\right)} \left(\theta _2^2+\theta _3^2\right) \left(\theta _2^{10}+\theta _3^{10}\right){}^2}{8 \sqrt{2} \eta ^{24}}\, .
\end{align}
The $q$-expansion of the defect Hilbert space partition function is, 
\begin{equation}
  \mathcal{Z}[D_{12}^2]_{\mathcal{N}}= q^{-15/16}+3q^{-7/16}+160q^{-1/16}+390q^{1/16}+160 q^{7/16}+4813 q^{9/16}+O\left(q^{15/16}\right)\, .
\end{equation}
The dimensions (spins) of the states in the defect Hilbert space are $s=\pm1/16+\mathbb{Z}/2$, which implies $\epsilon=+1$. Thus, we have the novel feature of having exchange outer automorphisms leading to the duality defect line $\mathcal{N}$. We also note the subtlety that $D_2$ should be treated in its $A_1\oplus A_1$ description. In the next section, we will present detailed results on self-dual orbifolds and defect partition functions for various $D_n$-type Niemeier lattice CFTs.

\section{Detailed results}
\label{sec:inner-outer-classification}

We will now provide detailed results on the orbifold partition functions for $D_n$-type Niemeier lattice CFTs for the non-reductive breakings. We provide results for the duality defect partition functions for $D_{24,1}$ and $D_{12,1}^2$. These capture all the essential features present in the other cases. For each duality defect, we categorise them into either the Ising or the $\text{Ising}\times SU(2)_1\text{WZW model}$ category.

\subsection{\ensuremath{\mathbf{D_{24,1}}}}

As discussed earlier, we consider two cases that lead to self-dual orbifolds: $(0,1)$ and $(4,0)$. For the other cases with non-anomalous $\mathbb{Z}_2$ leading to semi-simple factors, the orbifolded partition functions were already indicated in \eqref{D24-duality-web}. Let us now provide the detailed formulae for the duality defect partition functions in both the $(0,1)$ and the $(4,0)$ cases.

\subsubsection*{(0,1)}

The defect insertion partition function is computed similar to the calculation for the $E_{8,1}$ duality defect in \cite{Burbano:2021loy} and is given as,
\begin{equation}\label{d24-defect-insertionZ}
	\mathcal{Z}[D_{24}]^{\mathcal{N}_1}_i=\sqrt{2}\left(\chi_{0}^{so(48-i)}\chi_0^{so(i)}-\chi_{v}^{so(48-i)}\chi_v^{so(i)}\right)\, ,
\end{equation}
where, $i=1,3,5,\cdots,23$ and $i$ corresponds to cancelling the $i^{\text{th}}$ node in the $k=2$ twisted $D_{24}$ Dynkin diagram. The $q$-expansion of the defect Hilbert-space partition functions are tabulated in the table \ref{q-ser-d24-twistedZ} below.
\begin{table}[h]
\begin{center}
	\begin{tabular}{|C| L | L| }
		\hline
		\epsilon & \text{Subalgebra} & q\text{-expansion of }  	\mathcal{Z}^{D_{24}}_{\mathcal{N}_1}\\
		\hline
		+1 &\mathfrak{so}(47)\mathfrak{so}(1) & q^{-15/16}+47 q^{-7/16}+1082 q^{1/16}+16309q^{9/16}+O\left(q^{17/16}\right)\\
		\hline
		-1 & \mathfrak{so}(45)\mathfrak{so}(3) & 2q^{-13/16}+90q^{-5/16}+1986 q^{3/16}+28740 q^{11/16}+O\left(q^{17/16}\right)\\
		\hline
		-1 & \mathfrak{so}(43)\mathfrak{so}(5) & 4q^{-11/16}+172 q^{-3/16}+3632  q^{5/16}+50396 q^{13/16}+O\left(q^{17/16}\right)\\
		\hline
		+1 & \mathfrak{so}(41)\mathfrak{so}(7) & 8q^{-9/16}+328q^{-1/16}+6616 q^{7/16}+87904 q^{15/16}+O\left(q^{17/16}\right)\\
		\hline
		+1 & \mathfrak{so}(39)\mathfrak{so}(9)  & 16q^{-7/16}+624q^{1/16}+12000 q^{9/16}+O\left(q^{17/16}\right)\\
		\hline
		-1 & \mathfrak{so}(37)\mathfrak{so}(11) & 32q^{-5/16}+1184 q^{3/16}+21664 q^{11/16}+O\left(q^{17/16}\right)\\
		\hline
		-1 & \mathfrak{so}(35)\mathfrak{so}(13) & 64q^{-3/16}+2240 q^{5/16}+38912 q^{13/16}+O\left(q^{17/16}\right)\\
		\hline
		+1 & \mathfrak{so}(33)\mathfrak{so}(15)  & 128q^{-1/16}+4224 q^{7/16}+69504 q^{15/16}+O\left(q^{17/16}\right)\\
		\hline
		+1 & \mathfrak{so}(31)\mathfrak{so}(17) & 256 q^{1/16}+7936 q^{9/16}+32768 q^{15/16}+O\left(q^{17/16}\right)\\
		\hline
		-1 & \mathfrak{so}(29)\mathfrak{so}(19) & 512 q^{3/16}+14848 q^{11/16}+16384 q^{13/16}+O\left(q^{17/16}\right)\\
		\hline
		-1& \mathfrak{so}(27)\mathfrak{so}(21)  & 1024 q^{5/16}+8192 q^{11/16}+27648 q^{13/16}+O\left(q^{17/16}\right)\\
		\hline
		+1 & \mathfrak{so}(25)\mathfrak{so}(23) & 2048 q^{7/16}+4096 q^{9/16}+51200 q^{15/16}+O\left(q^{17/16}\right)\\
		\hline
	\end{tabular}
      \end{center}  
\caption{\label{q-ser-d24-twistedZ}The $q$-expansion for the twisted partition functions 
    obtained from the modular $S$-transformation of the partition functions in
    eqn.\eqref{d24-defect-insertionZ}. 
The $q$-expansion makes explicit the spins of the states in the twisted Hilbert space. Notice that there are an equal number of $\epsilon=+1$ and $\epsilon=-1$.}
\end{table}

As the inner automorphism $(0,1)$ did not come as a solution of the Ka\v c theorem, we did not review the lattice description in the previous section. In \cite{BoyleSmith:2023xkd}, it was shown that the projective phase for $(s)$ and $(c)$ conjugacy classes can also be implemented by taking $x \rightarrow x - \omega_1$. Note that $\omega_1 \in (v)$ and therefore $\omega_1\cdot \omega_1=1 \in 2\mathbb{Z}+1$. For the $(0,1)$ case , $-\omega_1$ is the new $x$ vector and therefore, from the lattice description $L^\prime = L_0 \cup ((L-L_0)+x)$ is the lattice corresponding to the orbifolded theory. This remains true for all the $D$-type Niemeier lattices. In the above case $L=(0)_{D_{24}}+(s)_{D_{24}}$ and $L_0=(0)_{D_{24}}, L-L_0=(s)_{D_{24}}$ therefore,

\begin{align}
 L_0 \cup ((L-L_0)+x) &= (0)_{D_{24}} \cup ((s)_{D_{24}}+(v)_{D_{24}})\nonumber\\
 &=(0)_{D_{24}} \cup (c)_{D_{24}}.
\end{align}

This also leads to the same partition function as the unorbifolded theory due to mapping under the exchange $(s)_{D_{24}}\leftrightarrow (c)_{D_{24}}$. Note that the $\mathbb{Z}_2$ duality defect of $E_{8,1}$ CFT can also be obtained from $(0)_{D_{8}}+(s)_{D_{8}}$ description in the same manner as above. 

\subsubsection*{(4,0)}
The defect insertion partition function can be written as,
\begin{align}\label{d24-defect-insertion_2}	\mathcal{Z}[D_{24}]^{\mathcal{N}_2}_i&=\sqrt{2}\chi_{0}^{so(40)}\left(\chi_{0}^{so(8-i)}\chi_0^{so(i)}-\chi_{v}^{so(8-i)}\chi_v^{so(i)}\right)\nonumber\\  &+\sqrt{2}\chi_{s}^{so(40)}\left(\chi_{v}^{so(8-i)}\chi_0^{so(i)}-\chi_{0}^{so(8-i)}\chi_v^{so(i)}\right)\, ,
\end{align}
where $i=1,3$.

The $q$-expansion of the defect partition functions are tabulated in the table \ref{q-ser-d24-twistedZ_2} below.
\begin{table}[h]
\begin{center}
	\begin{tabular}{|C| L | L |}
		\hline
		\epsilon & \text{Subalgebra} & q\text{-expansion of }  \mathcal{Z}^{D_{24}}_{\mathcal{N}_2}\\
		\hline
		+1 &\mathfrak{so}(40)\oplus\mathfrak{so}(7)\oplus\mathfrak{so}(1) &  
		q^{-15/16}+	7q^{-7/16}+	320q^{-1/16}+O(q^{1/16})\\
		\hline
		-1 &\mathfrak{so}(40)\oplus\mathfrak{so}(5)\oplus\mathfrak{so}(3) &  2q^{-13/16}+10q^{-5/16}+160q^{-3/16}+O(q^{3/16})\\
		\hline
	\end{tabular}
\caption{\label{q-ser-d24-twistedZ_2}The $q$-expansion for the twisted partition functions 
		obtained from the modular $S$-transformation of the partition functions in
		eqn.\eqref{d24-defect-insertion_2}. 
		The $q$-expansion makes explicit the spins of the states in the twisted Hilbert space. Notice that there are an equal number of $\epsilon=+1$ and $\epsilon=-1$.}
\end{center}
\end{table}

\subsection{\ensuremath{\mathbf{D_{12,1}^2}}}
In the $D_{12,1}^2$ there are three non-reductive self-dual orbifolds as discussed earlier. We will summarise the results for duality defect partition functions for each of these cases.

\subsubsection*{(0,0),(0,1)}
Recall that the glue code for $D_{12,1}^2$ CFT is,
\begin{equation}
	(0_{D_{12}},0_{D_{12}})+(c_{D_{12}},c_{D_{12}})+(s_{D_{12}},v_{D_{12}})+(v_{D_{12}},s_{D_{12}})\,. 
\end{equation}
If we have trivial automorphism $x$-vector for both the $D_{12,1}$, and a projective phase for the spinor and conjugate spinor of one of them, say the second, then the invariant sector is,
\begin{equation}
L_0=(0_{D_{12}},0_{D_{12}})+(s_{D_{12}},v_{D_{12}})\,. 
\end{equation}
To find the duality defect we need to find an outer autmorphism of the above lattice or the corresponding vertex operator algebra. The solution is clearly the automorphism $(s)_{D_{12}}\leftrightarrow (c)_{D_{12}}$ in the second $D_{12,1}$ factor. The corresponding duality defect partition functions are given by,
\begin{align}\label{d12-defect-insertionZ}	\mathcal{Z}[D_{12}^2]^{\mathcal{N}_1}_i&=\sqrt{2}\chi_{0}^{so(24)}\left(\chi_{0}^{so(24-i)}\chi_0^{so(i)}-\chi_{v}^{so(24-i)}\chi_v^{so(i)}\right)\nonumber\\ & +\sqrt{2}\chi_s^{so(24)}\left(\chi_{0}^{so(24-i)}\chi_v^{so(i)}-\chi_{v}^{so(24-i)}\chi_0^{so(i)}\right)\, ,
\end{align}
where $i=2n+1$, and $0\leq n\leq 5$, where different $i$ correspond to different phases added to the vertex operators by using Ka\v c theorem for the $k=2$ Dynkin diagram. The $q$-expansion of the partition functions for the defect Hilbert space are tabulated below in table \ref{q-ser-d12-twistedZ_2}.
\begin{table}[h]
\begin{center}
	\begin{tabular}{|C| L | L| }
		\hline
		\epsilon & \text{Subalgebra} & q\text{-expansion of }  \mathcal{Z}[D_{12}^2]_{\mathcal{N}_1}\\
		\hline
			
		+1 & \mathfrak{so}(24)\oplus\mathfrak{so}(23)\oplus\mathfrak{so}(1) &  24q^{-7/16}+552 q^{1/16}+2048 q^{7/16}+\mathcal{O}\left(q^{9/16}\right)\\
		\hline
		-1 & \mathfrak{so}(24)\oplus\mathfrak{so}(21)\oplus\mathfrak{so}(3) &  48q^{-5/16}+1008 q^{3/16}+1024 q^{5/16}+O\left(q^{11/16}\right)\\
			\hline
		-1 & \mathfrak{so}(24)\oplus\mathfrak{so}(19)\oplus\mathfrak{so}(5) &  96q^{-3/16}+512 q^{3/16}+1824 q^{5/16}+O\left(q^{11/16}\right)\\
			\hline
		+1 &\mathfrak{so}(24)\oplus\mathfrak{so}(17)\oplus\mathfrak{so}(7)  & 192q^{-1/16}+256q^{1/16}+3264 q^{7/16}+O\left(q^{9/16}\right)\\
			\hline
		+1 & \mathfrak{so}(24)\oplus\mathfrak{so}(15)\oplus\mathfrak{so}(9) &  128q^{-1/16}+384q^{1/16}+1152 q^{7/16}+O\left(q^{9/16}\right)\\
			\hline
		-1 & \mathfrak{so}(24)\oplus\mathfrak{so}(13)\oplus\mathfrak{so}(11)& 64q^{-3/16}+768 q^{3/16}+704 q^{5/16}+O\left(q^{11/16}\right)\\
		\hline
	\end{tabular}
\end{center}
\caption{\label{q-ser-d12-twistedZ_2}The $q$-expansion for the twisted partition functions 
		obtained from the modular $S$-transformation of the partition functions in
		eqn.\eqref{d12-defect-insertionZ}. 
		The $q$-expansion makes explicit the spins of the states in the twisted Hilbert space. Notice that there are an equal number of $\epsilon=+1$ and $\epsilon=-1$. }
\end{table}

\subsubsection*{(0,0),(4,0)}
We can write $D_{12,1}^2$ in the $D_4D_8D_{12}$ decomposition as,
\begin{align}
L=&(0_{D_{4}},0_{D_{8}},0_{D_{12}})+	(v_{D_{4}},v_{D_{8}},0_{D_{12}})+(c_{D_{4}},s_{D_{8}},c_{D_{12}})+(s_{D_{4}},c_{D_{8}},c_{D_{12}})\nonumber\\
	&+(s_{D_{4}},s_{D_{8}},v_{D_{12}})+(c_{D_{4}},c_{D_{8}},v_{D_{12}})+(0_{D_{4}},v_{D_{8}},s_{D_{12}})+(v_{D_{4}},0_{D_{8}},s_{D_{12}})\,. 
\end{align}
The duality defects are the automorphisms of the invariant sector,
\begin{align}
L_0&=(0_{D_{4}},0_{D_{8}},0_{D_{12}})+(s_{D_{4}},c_{D_{8}},c_{D_{12}})+(0_{D_{4}},v_{D_{8}},s_{D_{12}})+(s_{D_{4}},s_{D_{8}},v_{D_{12}})\,,
\end{align}
which is $(c)_{D_4}\leftrightarrow (v)_{D_4}$. The defect insertion partition function is given by,
\begin{align}\label{d12-defect-insertionZ2}
	\mathcal{Z}[D_{12}^2]^{\mathcal{N}_2}_i&=\sqrt{2}\left(\chi_0^{so(i)}\chi_0^{so(8-i)}-\chi_v^{so(i)}\chi_v^{so(8-i)}\right)\left(\chi_0^{D_8}\chi_0^{D_{12}}+\chi_v^{D_8}\chi_s^{D_{12}}\right)\nonumber\\
	&\quad+\sqrt{2}\left(\chi_0^{so(8-i)}\chi_v^{so(i)}-\chi_v^{so(8-i)}\chi_0^{so(i)}\right)\left(\chi_c^{D_8}\chi_c^{D_{12}}+\chi_c^{D_8}\chi_v^{D_{12}}\right)\, ,
\end{align}
where $i=1,3$.

\begin{table}[h]
\begin{center}
	\begin{tabular}{|C| L | L |}
		\hline
		\epsilon & \text{Subalgebra} & q\text{-expansion of }  \mathcal{Z}[D_{12}^2]_{\mathcal{N}_2}\\
		\hline
		+1 & \mathfrak{so}(24)\oplus\mathfrak{so}(16)\oplus \mathfrak{so}(7)\oplus\mathfrak{so}(1)  & 8q^{-9/16}+16q^{-7/16}+8q^{-1/16}+O\left(q^{1/16}\right)\\
		\hline
		-1 & \mathfrak{so}(24)\oplus\mathfrak{so}(16)\oplus \mathfrak{so}(5)\oplus\mathfrak{so}(3) & 4q^{-11/16}+32q^{-5/16} +12q^{-3/16}+O(q^{3/16})\\
		\hline
	\end{tabular}
\end{center}
\caption{The $q$-expansion for the twisted partition functions 
		obtained from the modular $S$-transformation of the partition functions in
		eqn.\eqref{d12-defect-insertionZ2}. 
		The $q$-expansion makes explicit the spins of the states in the twisted Hilbert space. Notice that there are an equal number of $\epsilon=+1$ and $\epsilon=-1$.}
\end{table}

\subsection{\ensuremath{\mathbf D_{8,1}^3}}
The independent non-anomalous discrete symmetries of $D_{8,1}^3$ CFT and the corresponding orbifold theories we obtain from Ka\v c's theorem are given in the table \ref{orbifolds_d8^3}. Note that we only consider the orbifolds, which give us a semi-simple current algebra.
\begin{table}[h]
\begin{center}
	\begin{tabular}{|L|L|C|}
		\hline		
		i&\text{Automorphism}& \text{Algebra for } \mathcal{Z}^{\text{orb}}\\
		\hline
		\rowcolor{lightcyan}
		1&(0,0),(0,0),(0,1)& D_8^3\\
		\hline
		\rowcolor{lightcyan}
		2&(0,0),(0,0),(4,0)& D_8^3\\
		\hline
		3&(0,0), (0,1), (0,1)& D_{16}E_8\\
		\hline
		4&(0,0), (0,1), (4,0)&D_{12}^2\\
		\hline
		\rowcolor{lightcyan}
		5&(0,0), (2,0), (2,0)&D_8^3\\
		\hline
		\rowcolor{lightcyan}
		6&(0,0), (4,0), (4,0)&D_8^3\\
		\hline
		7&(0,1), (0,1), (0,1)&E_8^3\\
		\hline
		\rowcolor{lightcyan}
		8&(0,1), (0,1), (4,0)&D_8^3\\
		\hline
		9&(0,1), (2,0), (2,0)&D_{10}E_7^2\\
		\hline
		\rowcolor{lightcyan}
		10&(0,1), (4,0), (4,0)&D_8^3\\
		\hline
		11&(2,0), (2,0), (4,0)& D_6^4\\
		\hline 
		12&	(4,0), (4,0), (4,0)& D_4^6\\
		\hline
	\end{tabular}
\end{center} 
\caption{\label{orbifolds_d8^3} The non-anomalous orbifolds of the $D_{8}^3$ Niemeier lattice CFT obtained from Ka\v c's theorem. The self-dual orbifolds are marked in blue. }
\end{table}

The self-dual orbifolds of $D_{8,1}^3$ meromorphic CFT and the outer automorphisms of the invariant sector under the corresponding orbifolding are tabulated in the table \ref{d8^3_dualitydefect_table}. In the tables, we use the symbols given in table \ref{symbol_outer} to denote the Dynkin diagram symmetries of the corresponding algebra.
\begin{table}
\begin{center}
	\begin{tabular}{|l|l|}
		\hline
		Symbol & Diagram automorphism\\
		\hline
		$\mathcal{I}$ & \text{Identity}\\
		\hline
		$\mathcal P_v$  & $(s)_{D_{r}}\leftrightarrow(c)_{D_{r}}\, \forall r$\\
		\hline
		$\mathcal P_s$  & $(v)_{D_{4}}\leftrightarrow(c)_{D_{4}}$\\
		\hline
		$\mathcal P_c$  & $(v)_{D_{4}}\leftrightarrow(s)_{D_{4}}$\\
		\hline
		$\mathcal T$	& $(c)_{D_{4}}\rightarrow(v)_{D_{4}}\rightarrow(s)_{D_{4}}\rightarrow(c)_{D_{4}}$\\
		\hline
		$\mathcal T^2$	& $(v)_{D_{4}}\rightarrow(c)_{D_{4}}\rightarrow(s)_{D_{4}}\rightarrow(v)_{D_{4}}$\\
		\hline
	\end{tabular}
\end{center}
\caption{\label{symbol_outer} The symbols for the symmetries of $D_n$ Dynkin diagrams.}
\end{table}
Apart from the outer automorphisms of the invariant sector given by Dynkin diagram symmetries, we also encounter the \textit{exchange symmetries}. The exchanges are written explicitly in the table \ref{d8^3_dualitydefect_table}. For each self-dual orbifold case, duality defect partition functions can be calculated using the same techniques as demonstrated in section \ref{sec:D-def-key}.
\begin{table}
\begin{center}
	\begin{tabular}{|L| L| L |L |}
		\hline
		i& \text{Symmetry of CFT} & \text{Fixed subalgebra} & \text{Outer transformations} \\
		\hline
		1& (0,0),(0,0),(0,1)& D_8^{(1)}D_8^{(2)}D_8^{(3)} &  
		\mathcal P_{v,{D_8^{(3)}}}\\[3mm]
		\hline
		2 &(0,0),(0,0),(4,0) & D_8^{(1)}D_8^{(2)}D_4^{(1)}D_4^{(2)} &\mathcal P_{s,{D_4^{(2)}}} \\
		[3mm]\hline	
		3 &(0,0),(2,0),(2,0) & D_8D_6^{(1)}A_1^{(1)}A_1^{(2)}D_6^{(2)}A_1^{(3)}A_1^{(4)}& A_1^{(2)}\leftrightarrow A_1^{(4)}\\
		[3mm]\hline
          4&(0,0),(4,0),(4,0) & D_8 D_4^{(1)}D_4^{(2)}D_4^{(3)}D_4^{(4)} & D_4^{(2)}\leftrightarrow D_4^{(3)},\mathcal P_{v,D_8}, \mathcal P_{s,D_4^{(1)}},\\
                 &&&\mathcal T_{D_4^{(2)}},\mathcal T^2_{D_4^{(3)}},\mathcal P_{c,D_4^{(4)}}\\
          \hline
          5 & (0,1), (0,1),(4,0) & D_8^{(1)}D_8^{(2)}D_4^{(1)}D_4^{(2)} &  \mathcal P_{s,{D_4^{(1)}}}\\
          [3mm]\hline
          6 & (0,1),(4,0),(4,0) & D_8 D_4^{(1)}D_4^{(2)}D_4^{(3)}D_4^{(4)} & \, D_4^{(2)}\leftrightarrow D_4^{(4)},\mathcal P_{v,D_8},\mathcal P_{c,D_4^{(1)}},\\
                 &&&\mathcal P_{s,D_4^{(2)}},\mathcal P_{c,D_4^{(3)}},\mathcal P_{s,D_4^{(4)}}\\
		\hline
	\end{tabular}
\end{center}
\caption{\label{d8^3_dualitydefect_table} The outer automorphisms of the untwisted sector under orbifolding of the $D_8^3$ Niemeier lattice CFT with the corresponding fixed subalgebras.}
\end{table}
The counting of automorphisms of the invariant lattice under orbifolding can be tricky. We encounter cases where several seemingly different outer automorphisms are found to have an identical action. In other words, identical automorphisms can be realised in different ways, which we carefully exclude to avoid double counting. For example, in the case of the orbifold symmetry $(0,0),(4,0),(4,0)$ of the $D_8^3$, CFT we find that the automorphism $D_{4}^{(1)}\rightarrow D_{4}^{(2)}\rightarrow D_{4}^{(4)}\rightarrow D_{4}^{(1)},\mathcal P_{v,D_8},\mathcal T^2_{D_4^{(1)}},\mathcal P_{c,D_4^{(2)}},\mathcal P_{s,D_4^{(3)}},\mathcal T_{D_4^{(4)}}$ which is a combination of the $\mathbb{Z}_3$ exchange, and $\mathbb{Z}_2\, ,\mathbb{Z}_3$ symmetries of the $D_8$ and $D_4$ factors is identical to  $D_4^{(2)}\leftrightarrow D_4^{(3)},\mathcal P_{v,D_8}, \mathcal P_{s,D_4^{(1)}},
\mathcal T_{D_4^{(2)}},\mathcal T^2_{D_4^{(3)}},\mathcal P_{c,D_4^{(4)}}$ given in the table \ref{d8^3_dualitydefect_table}. 
Further, in some cases, the outer automorphism of the invariant sublattice is also a symmetry of the full lattice. These are the symmetries of the orbifold theory and correspond to invertible lines. These cases are also excluded from the classification of duality defects, which are non-invertible in the unorbifolded theory, as opposed to symmetries of the full lattice, which are invertible. An example of such a symmetry is the $D_4^{(1)}\leftrightarrow D_4^{(2)},D_4^{(3)}\leftrightarrow D_4^{(4)}, \mathcal P_{v,D_4^{(1)}},\mathcal P_{v,D_4^{(2)}}, \mathcal P_{v,D_4^{(3)}},\mathcal P_{v,D_4^{(4)}}$. These are typically easier to identify due to their close relation with the symmetry of the original unorbifolded theory.

Another interesting defect is the operation $D_4^{(2)}\leftrightarrow D_4^{(3)},\mathcal P_{v,D_8}, \mathcal P_{s,D_4^{(1)}},\mathcal T_{D_4^{(2)}},\mathcal T^2_{D_4^{(3)}},\mathcal P_{c,D_4^{(4)}}$ in table \ref{d8^3_dualitydefect_table} for $i=4$. Note that this is not a triality defect even though we have a $\mathbb{Z}_3$ action and another factor with a $\mathbb{Z}_2$ action. Pure algebraically, this is a member of the $\mathbb{Z}_6$ group, but the squared action on the twisted Hilbert space (under orbifolding) gives back the twisted Hilbert space. The non-invariant part of the glue code and the twisted sector are mapped to each other under the outer automorphism, which implies the above defect is a duality defect.


The tables show that there always exists a unique defect operator for each orbifold symmetry. The existence of the unique defect operator is guaranteed since the unique outer automorphism always exists, which takes the glue code of the original CFT to the glue code of the orbifolded CFT.

\subsection{\ensuremath{\mathbf{D_{6,1}^4}}}

Similarly, for the CFT with current algebra $D_{6,1}^4$, we find the following non-anomalous independent orbifolds with a semi-simple current algebra noted in table \ref{orbifolds_d6^4}.
\begin{table}[h]
\begin{center}
	\begin{tabular}{|L|L|C|}
		\hline		
		i&\text{Automorphism}& \text{Algebra for } \mathcal{Z}^{\text{orb}}\\
		\hline
		\rowcolor{lightcyan}
		1&(0,0),(0,0),(0,0),(0,1) & D_6^4\\
		\hline
		2&(0,0),(0,0),(0,1),(0,1) & D_{12}^2\\
		\hline
		\rowcolor{lightcyan}
		3&(0,0),(0,0),(2,0),(2,0) & D_6^4\\
		\hline
		\rowcolor{lightcyan}
		4&(0,0),(0,1),(0,1),(0,1) &D_6^4\\
		\hline
		5&(0,0),(0,1),(2,0),(2,0) &D_8^3\\
		\hline
		\rowcolor{lightcyan}
		\hline
		\rowcolor{lightcyan}
		6&(0,1),(0,1),(2,0),(2,0) &D_6^4\\
		\hline
		7&(2,0),(2,0),(2,0),(2,0) &D_4^6\\
		\hline
		8&(3,0),(3,0),(3,0),(3,0) &A_3^8\\
		\hline
	\end{tabular}
\end{center}
\caption{\label{orbifolds_d6^4}The non-anomalous orbifolds of the $D_{6}^4$ Niemeier lattice CFT obtained from Ka\v c's theorem. The self-dual orbifolds are marked in blue.}
\end{table}

We list the self-dual orbifolds of $D_{6,1}^4$ Niemeier lattice CFT with the corresponding independent outer automorphism of the untwisted sector. The results are tabulated in table \ref{d6^4_dualitydefect_table}. 
\begin{table}[h]
\begin{center}
	\begin{tabular}{|L| L| L|L |}
		\hline
		i& \text{Symmetry of CFT} & \text{Fixed subalgebra} & \text{Outer transformations} \\
		\hline
		1& (0,0),(0,0),(0,0),(0,1) & D_6^{(1)}D_6^{(2)}D_6^{(3)}D_6^{(4)}& \mathcal P_{v,D_6^{(4)}}\\
		\hline
		2& (0,0),(0,0),(2,0),(2,0)& D_6^{(1)} D_6^{(2)} D_4^{(1)} A_1^{(1)}A_1^{(2)}D_4^{(2)} A_1^{(3)}A_1^{(4)}& A_1^{(2)}\leftrightarrow A_1^{(4)}\\
		
		\hline
		3& (0,0),(0,1),(0,1),(0,1) & D_6^{(1)}D_6^{(2)}D_6^{(3)}D_6^{(4)}&  \mathcal P_{v,D_6^{(1)}}\\
		\hline
		4& (0,1),(0,1),(2,0),(2,0)&D_6^{(1)} D_6^{(2)} D_4^{(1)} A_1^{(1)}A_1^{(2)}D_4^{(2)} A_1^{(3)}A_1^{(4)}&  A_1^{(1)}\leftrightarrow A_1^{(3)}\\
		
		\hline
	\end{tabular}
\end{center}
\caption{\label{d6^4_dualitydefect_table} The outer automorphisms of the untwisted sector under orbifolding of the $D_6^4$ Niemeier lattice CFT with the corresponding fixed subalgebras.}
\end{table}
There exist an automorphism $(0,1),(0,1),(0,1),(0,1)$ which looks like a $\mathbb{Z}_2$ symmetry but this keeps the full glue code invariant. In terms of the lattice vector addition rule, notice that $x=(v,v,v,v)\in L[D_6^4]$, so the lattice remains invariant under orbifolding. This implies that $(0,1),(0,1),(0,1),(0,1)$ is the identity operator on the $D_6^4$ CFT. We also encounter another such automorphism in $D_4^6$ CFT.

\subsection{\ensuremath{\mathbf{D_{4,1}^6}}}

For the CFT with current algebra $D_{4,1}^6$, we only give the non-anomalous independent orbifolds obtained from Ka\v c's theorem with a semi-simple current algebra in table \ref{orbifolds_d4^6}.
\begin{table}[h]
\begin{center}
	\begin{tabular}{|L|L|C|}
		\hline		
		i&\text{Automorphism}& \text{Algebra for } \mathcal{Z}^{\text{orb}}\\
		\hline
		\rowcolor{lightcyan}
		1&(0,0),(0,0),(0,0),(0,0),(0,0),(0,1) & D_4^6\\
		\hline
		2&(0,0),(0,0),(0,0),(0,0),(0,1),(0,1) & D_8^3\\
		\hline
		\rowcolor{lightcyan}
		3&(0,0),(0,0),(0,0),(0,0),(2,0),(2,0) & D_4^6\\
		\hline
		4&(0,0),(0,0),(0,0),(0,1),(0,1),(0,1) & D_8^3\\
		\hline
		5&(0,0),(0,0),(0,0),(0,1),(2,0),(2,0) & D_6^4 \\
		\hline
		6&(0,0),(0,0),(0,1),(0,1),(0,1),(0,1) & D_8^3\\
		\hline
		7&(0,0),(0,0),(0,1),(0,1),(2,0),(2,0) & D_6^4\\
		\hline
		\rowcolor{lightcyan}
		8&(0,0),(0,0),(2,0),(2,0),(2,0),(2,0) & D_4^6\\
		\hline
		\rowcolor{lightcyan}
		9&(0,0),(0,1),(0,1),(0,1),(0,1),(0,1) & D_4^6\\
		\hline
		10&(0,0),(0,1),(0,1),(0,1),(2,0),(2,0) & D_6^4\\
		\hline
		\rowcolor{lightcyan}
		11&(0,0),(0,1),(2,0),(2,0),(2,0),(2,0) & D_4^6\\
		\hline
		\rowcolor{lightcyan}
		\hline
		\rowcolor{lightcyan}
		12&(0,1),(0,1),(0,1),(0,1),(2,0),(2,0) &D_4^6\\
		\hline
		\rowcolor{lightcyan}
		13&(0,1),(0,1),(2,0),(2,0),(2,0),(2,0) &D_4^6\\
		\hline
		14&(2,0),(2,0),(2,0),(2,0),(2,0),(2,0) &A_1^{24}\\
		\hline
	\end{tabular}
\end{center}
\caption{\label{orbifolds_d4^6}The non-anomalous orbifolds of the $D_{4}^6$ Niemeier lattice CFT obtained from Ka\v c's theorem. The self-dual orbifolds are marked in blue.}
\end{table}

We find the automorphism $(0,1),(0,1),(0,1),(0,1),(0,1),(0,1)$ which looks like a $\mathbb{Z}_2$ symmetry, but is again, the identity operator. 

Calculating the outer automorphisms corresponding to duality defects and corresponding partition functions is analogous to previous cases. The computations for these cases quickly become huge but are straightforward. Note that since the invariant sub-algebras have many factors of $D_4$,  it may lead to many cases with $\mathbb{Z}_3$ automorphisms of factors, but the defects are duality defects, like the ones we encountered above.

\section{Discussion}\label{sec:discussion}

We discussed duality defects in the $c=24$ meromorphic CFTs by focusing on the Niemeier Lattice CFTs corresponding to $D_{n}$-type lattices. We classified anomaly-free $\mathbb{Z}_{2}$ orbifolds of $D_{24,1}$, $D_{4,1}\times D_{20,1}$, $(D_{12,1})^{2}$, $(D_{8,1})^{3}$, $(D_{6,1})^{4}$, and $(D_{4,1})^{6}$ groups based on inner and outer automorphisms and studied duality defects corresponding to these automorphisms. This analysis is based on the Ka\v c theorem, which classifies the automorphisms of the groups. Using the inner automorphisms, we showed that one can recover the self-dual lattice by combining the invariant and appropriately twisted lattice corresponding to the $\mathbb{Z}_{2}$ orbifold.

Besides these duality defects, we also explored the outer automorphisms in the cases where the groups repeated, e.g., $(D_{12,1})^{2}$, where a delicate choice of the exchange automorphism led to defect partition function symmetric under this exchange symmetry. The simplest example of this exchange automorphism occurs in $(D_{12,1})^{2}$ case, and we explicitly worked it out in this case. A similar exchange exists for $(D_{8,1})^{3}$, $(D_{6,1})^{4}$, and $(D_{4,1})^{6}$ cases with the permutation group symmetries $S_{3}$, $S_{4}$, and $S_{6}$ respectively.  While the extension of the analysis of exchange automorphism to these cases is straightforward, the explicit computation quickly becomes tedious.

The meromorphic CFTs are relevant for the $N=(0,2)$ heterotic string theory\cite{Ooguri:1991ie}, which has $N=2$ superstring in the right chiral sector and bosonic string on the left. Since the chiral critical dimension of this theory is 2, we need to compactify the bosonic string theory on a 24-dimensional even self-dual lattice, which is precisely the Schellekens list\cite{Schellekens:1992db}. It would be interesting to explore the utility of duality defects in the $(0,2)$ heterotic string.

In principle, our procedure can also be extended to study other Niemeier lattice theories. It would be interesting to classify the defect partition functions in these cases as well as those meromorphic theories which are beyond the Niemeier CFTs. In particular, it would be interesting to see how the exchange automorphism plays out in these cases. We hope to return to these problems soon. Finally, the $\mathbb{Z}_{n}$ automorphisms for $n>2$ will lead to a more intricate structure of defects, and it would be interesting to investigate them.


\section*{Acknowledgements}
We thank Sujay K. Ashok and Sunil Mukhi for enlightening discussions. We thank the organisers of \emph{Indian Strings Meeting 2023} at the Indian Institute of Technology Bombay for facilitating discussions. SG acknowledges the generous support of the Infosys Foundation. SG thanks the Institute of Mathematical Sciences, Chennai, for hospitality. SH thanks Semanti Dutta and Arnab Priya Saha for their hospitality at the Indian Institute of Science. SH thanks Harish-Chandra Research Institute, Allahabad, for hospitality.

\appendix
\section{Useful formulae}\label{app:formulae}
Here we provide the characters of $D_{n,1}$ type theories in terms of theta functions. We also give the well known theta function identities including its modular properties. In the second part of this appendix we provide the decomposition of the conjugacy classes of $D_n$ theories in terms of smaller $D_n$ algebras.

\subsection{\ensuremath{D_n}-current algebras and Jacobi theta functions}
There are four conjugacy classes in the $D$-type Lie algebras, which are the vector, spinor and conjugate spinor conjugacy classes denoted by $(v)$, $(s)$ and $(c)$ notation in the main text. The $q$-characters of the $D_n$-type algebras can be written in terms of Jacobi theta functions,
\begin{align}
	\chi_{0}^{D_n}(q)&=\frac{\theta_3^{n}(q)+\theta_4^{n}(q)}{2\eta^n(q)}\, ,\nonumber\\
	\chi_{v}^{D_n}(q)&=\frac{\theta_3^{n}(q)-\theta_4^{n}(q)}{2\eta^n(q)}\, ,\nonumber\\
	\chi_{s}^{D_n}(q)&=\chi_{c}^{D_n}(q)=\frac{\theta_2^{n}(q)}{2\eta^n(q)}\, ,
\end{align}
where $q=e^{2\pi i\tau}$, $\tau\in\mathbb{H}$. These expansions are used to obtain the $q$-expansion of various partition functions.

The modular transformation of these characters is easily obtained from the lattice description (see for example \cite{Lerche:1988np}),
\begin{align}
	\chi_{0}^{D_n}(\tilde q)&=\frac{1}{2}\left(\chi_0^{D_n}(q)+\chi_v^{D_n}(q)+\chi_s^{D_n}(q)+\chi_c^{D_n}(q)\right)\, ,\nonumber\\
	\chi_{v}^{D_n}(\tilde q)&=\frac{1}{2}\left(\chi_0^{D_n}(q)+\chi_v^{D_n}(q)-\chi_s^{D_n}(q)-\chi_c^{D_n}(q)\right)\, ,\nonumber\\
	\chi_{s}^{D_n}(\tilde q)&=\frac{1}{2}\left(\chi_0^{D_n}(q)-\chi_v^{D_n}(q)+\chi_s^{D_n}(q)-\chi_c^{D_n}(q)\right)\, ,\nonumber\\
	\chi_{c}^{D_n}(\tilde q)&=\frac{1}{2}\left(\chi_0^{D_n}(q)-\chi_v^{D_n}(q)-\chi_s^{D_n}(q)+\chi_c^{D_n}(q)\right)\, , \nonumber
\end{align}
where $\tilde q=e^{-2\pi i/\tau}$.

The definition of $\theta(\tau)$ functions used in the main text is given below \cite{DiFrancesco:1997nk}\footnote{also look at \cite{Dabholkar:2005dt}}
\begin{align}\label{theta-def}
	\theta_1(2\tau)&=\sum_{n\in \mathbb{Z}} q^{(n+1/2)^2}(-1)^n=\sum_{n\in \mathbb{Z}} q^{(-n+1/2)^2}(-1)^n=\sum_{n\in \mathbb{Z}} q^{(n+1/2)^2}(-1)^{n+1}=0\ ,\nonumber\\
	\theta_2(2\tau)&=\sum_{n\in \mathbb{Z}} q^{(n+1/2)^2},\nonumber\\
	\theta_3(2\tau)&=\sum_{n\in \mathbb{Z}} q^{n^2}\ ,\nonumber\\
	\theta_4(2\tau)&=\sum_{n\in \mathbb{Z}} q^{n^2}(-1)^n \ .
\end{align}
For $\theta_1$ we transform the summation variable $n\rightarrow -n$ to get the second equality and $n\rightarrow n+1$ and the fact that $(-n+1/2)^2=(n-1/2)^2$ to get the third equality. Since we get an extra negative sign in going from first to last equality, $\theta_1=0$.

Note the modular transformation of the $\theta(\tau)$ functions and the Dedekind $\eta(\tau)$ function is \cite{DiFrancesco:1997nk},
\begin{equation}\label{s-t-theta}
	\begin{matrix}
		\theta_2(-1/\tau)=\sqrt{-i\tau}\theta_4(\tau)\ , & \theta_2(\tau+1)=e^{i\pi/4}\theta_2(\tau)\ , \nonumber\\
		\theta_3(-1/\tau)=\sqrt{-i\tau}\theta_3(\tau)\ , &
		\theta_3(\tau+1)=\theta_4(\tau)\ ,  \nonumber\\
		\theta_4(-1/\tau)=\sqrt{-i\tau}\theta_2(\tau)\ , & 
		\theta_4(\tau+1)=\theta_3(\tau)\ ,  \nonumber\\
		\eta(-1/\tau)=\sqrt{-i\tau}\eta(\tau)\ , & \eta(\tau+1)=e^{i\pi/12}\eta(\tau) .
	\end{matrix}
\end{equation}

\subsection{\ensuremath{D_r\times D_{n-r}} subalgebra inside \ensuremath{D_{n}}}\label{d-to-d-decomposition}
As discussed earlier, $(r,0)$ inner automorphism of $D_{n,1}$ leads to the decomposition $D_{n,1} \rightarrow D_{r,1}\times D_{n-r,1}$. We demonstrate this with the example of $D_{24}$. Let us see below how the original lattice $(0)_{D_{24}}+(s)_{D_{24}}$ decomposes in this description.

Constructing $(0)_{D_{24}}+(s)_{D_{24}}$ from $D_r\times D_{24-r}$: The first step in developing an understanding is to reconstruct the original lattice $L$ from the fixed lattice $L_0$. This is seen from the counting of dimension of the irreducible representations of the $D_{24}$ finite Lie algebra which are lifted to the representations of the affine Lie algebra $D_{24}$ at level 1.

The dimension of $D_{24}$ root lattice subtracting the Cartans is $24\times47-24=1128-24$ (or the number of norm $2$ vectors which are the number of roots of $D_{24}$) and the conjugacy class $(s)_{D_{24}}$ has a Dirac spinor of dimension $2^{24}$ which is reducible to the irreducible representation of Weyl spinor of dimension $2^{23}$. To construct the adjoint and spinor representation of $D_{24}$ from the representations of $D_r\times D_{24-r}$ we note that the following dimension of irreducible representations of the a $D_r$ algebra:
\begin{align}\label{dimensions_of_D_r}
	\text{dim } (0)_{D_{r}}&=r(2r-1)-r\, ,\nonumber\\
	\text{dim } (s)_{D_{r}}&=\text{dim } (c)_{D_{r}}=2^{r-1}\, ,\nonumber\\
	\text{dim } (v)_{D_{r}}&=2r\, .
\end{align}
The conjugacy classes of $D_r\times D_{24-r}$,  $(0_{D_{r}},0_{D_{24-r}})+(v_{D_{r}},v_{D_{24-r}})$ have the total dimension $1128-24$ which belongs to the adjoint representation of $D_{24}$. The vector conjugacy class of $D_{24}$ can be decomposed as $(0_{D_{r}},v_{D_{24-r}})+(v_{D_{r}},0_{D_{24-r}})$ of $D_r\times D_{24-r}$ 

The conjugacy $(s)_{D_{24}}$ is constructed from the representation $(s_{D_{r}},s_{D_{24-r}})+(c_{D_{r}},c_{D_{24-r}})$. To check this first note that the dimensions of the representations match \eqref{dimensions_of_D_r} which gives a spinor (conjugate) representation. The eigenvalue of the chirality operator ($+1$ for spinor and $-1$ for conjugate spinor, say) provides a definite test of the nature of the spinor in a purely coordinate independent way.

\begin{equation}\label{chirality_operator_in_48_dim }
	\Gamma_{(48)}=i^{-23}\Gamma^0\Gamma^1\cdots\Gamma^{47}\, ,
\end{equation}
decomposes as $\Gamma_{(2r)}\otimes\Gamma_{(48-2r)}$ since,
\begin{align}
	\left\lbrace \Gamma_i, \Gamma_j \right\rbrace&=2\eta_{ij}\, i,j=0,1,\cdots, 2r-1\\
	\left\lbrace\Gamma_I, \Gamma_J \right\rbrace&=2\delta_{IJ}\, I,J=2r,1,\cdots, 47
\end{align}
which can give the correct metric on the right hand side of the second equation above if we make the change $\Gamma_{2r}\to i\Gamma_{2r}$, resulting in,
\begin{equation}
	\left\lbrace\Gamma_I, \Gamma_J\right\rbrace=2\eta_{IJ}\, I,J=2r,1,\cdots, 47
\end{equation}
Thus, $(s)_{D_{24}}=(s_{D_{r}},s_{D_{24-r}})+(c_{D_{r}},c_{D_{24-r}})$ has the correct eigenvalue of $\Gamma_{(48)}$, $+1$. The fact that the conjugacy classes $(s_{D_{r}},s_{D_{24-r}})+(c_{D_{r}},c_{D_{24-r}})$ are a representation of $(s)_{D_{24}}$ instead of $((s_{D_{r}},c_{D_{24-r}})+(c_{D_{r}},s_{D_{24-r}})$ can also be checked from a particular coordinate representation such as the one in \cite{Lerche:1988np}. In other words, the spinor is written as,
\begin{equation}\label{spinor_decomposition}
	\xi^{(s)}_{24}=\xi^{(s)}_{r}\otimes\xi^{(s)}_{24-r}+\xi^{(c)}_{r}\otimes\xi^{(c)}_{24-r}\, ,
\end{equation}
and the conjugate spinor is,
\begin{equation}
	\xi^{(c)}_{24}=\xi^{(s)}_{r}\otimes\xi^{(c)}_{24-r}+\xi^{(c)}_{r}\otimes\xi^{(s)}_{24-r}\, .
\end{equation}

\section{Defect insertion partition function letters}

We list the `letters' composing the defect insertion partition function corresponding to the outer automorphisms listed in tabel \ref{symbol_outer}.

\begin{enumerate}
	\item $\mathcal{P}_{s,D_4^{(i)}}$, $\mathcal{P}_{c,D_4^{(i)}}$ which has the action $v\leftrightarrow c$ and $v\leftrightarrow s$ on the $i$-th $D_4$ factor of the fixed subalgebra under orbifold. We have already seen the action of $\mathcal{P}_{s,D_4^{(i)}}$ while calculating the lattice sum in \eqref{Z1-D24-insertion}, for example. The contribution of each part is given by,
	\begin{align}
		&\chi_0^{so(8-i)}\chi_0^{so(i)}-\chi_v^{so(8-i)}\chi_v^{so(i)}\, , \text{ and} \\
		&\chi_0^{so(8-i)}\chi_v^{so(i)}-\chi_v^{so(8-i)}\chi_0^{so(i)}\, ,
	\end{align}
respectively, which needs to be multiplied by appropriate letters of the remaining lattice contributing to the partition function.

	\item $\mathcal{P}_{v,D_n^{(i)}}$,  which has the action $s\leftrightarrow c$ on the $i$-th $D_n$ factor of the fixed subalgebra under orbifold, where $n\in\mathbb{Z}_{\geq2}$. This automorphism keeps fixed the $C_{n-1}$ lattice and the $v$ conjugacy class of the invariant Hilbert space under orbifolding. The contribution is exactly the same as above,
	\begin{align}
		&\chi_0^{so(2n-i)}\chi_0^{so(i)}-\chi_v^{so(2n-i)}\chi_v^{so(i)}\, , \text{ and} \\
		&\chi_0^{so(2n-i)}\chi_v^{so(i)}-\chi_v^{so(2n-i)}\chi_0^{so(i)}\, .
	\end{align}
	This is because, again we have $\omega_1^{D_n}=\omega_1^{C_{n-1}}$ composing of the linear combination of roots which are held fixed under the outer automorphism.

	\item $\mathcal{T}$, and $\mathcal{T}^2$  which are $\mathbb{Z}_3$ symmetries the of $D_4$ Dynkin diagram, have the action $(c)_{D_{4}}\rightarrow(v)_{D_{4}}\rightarrow(s)_{D_{4}}\rightarrow(c)_{D_{4}}$ and $(v)_{D_{4}}\rightarrow(c)_{D_{4}}\rightarrow(s)_{D_{4}}\rightarrow(v)_{D_{4}}$ respectively. From the Dynkin diagram of $D_4$, 
	
	
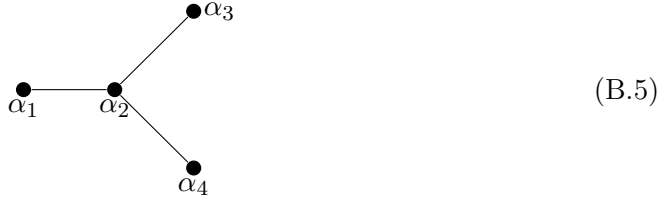
\begin{equation}
		\begin{tikzpicture}[baseline={(current bounding box.center)}, scale = 1]
			\tikzstyle{vertex}=[circle, fill=black, minimum size=2pt,inner sep=2pt];
			\def\r{1.2};
			\node[vertex] (T1) at (\r*5,\r*0) {};
			\node[vertex] (T2) at (\r*6,\r*0) {};
			\node[vertex] (T3) at (\r*6.866,\r*0.866) {};
			\node[vertex] (T4) at (\r*6.866,-1*\r*0.866) {};
			
			\draw[-] (T1) -- (T2);
			\draw[-] (T2) -- (T3);
			\draw[-] (T2) -- (T4);
			\draw[below] (T1) node {$\alpha_1$};
			\draw[below] (T2) node {$\alpha_2$};
			\draw[right] (T3) node {$\alpha_3$};
			\draw[below] (T4) node {$\alpha_4$};
		\end{tikzpicture}
		\label{d4v-d4c}
	\end{equation}

	it is clear that the invariant lattice is,
	\begin{equation}
		L_0=\mathbb{Z}\alpha_2+\mathbb{Z}(\alpha_1+\alpha_3+\alpha_4)\, .
	\end{equation}
	The two simple roots of the algebra generating the lattice are $\alpha_2$ (short root) and $\alpha_1+\alpha_3+\alpha_4$ (long root). These are the roots of the $G_2$ algebra. The $G_2$ root lattice is isomorphic to the $A_2$ root lattice. Thus the lattice sum can be performed over the $A_2$ root lattice. The lattice sum is given in \cite{Conway:1988oqe}. 
	
	The oscillator contribution can be calculated to give,
	\begin{equation}
		\eta_g(\tau)=\eta(\tau)\eta(3\tau)\, .
	\end{equation}

	Thus, the total contribution is,
	\begin{align}
		\frac{1}{\eta(\tau)\eta(3\tau)}\sum_{\alpha\in\Lambda_{A_2}}q^{(\alpha,\alpha)/2}&=\frac{1}{\eta(\tau)\eta(3\tau)}\sum_{x_1,\, x_2\in\mathbb{Z}}q^{x_1^2+x_2^2-x_1x_2}\, ,\nonumber\\
		&=\frac{1}{\eta(\tau)\eta(3\tau)}
		\left( \theta_3(2\tau)\theta_3(6\tau)+\theta_2(2\tau)\theta_2(6\tau)\right)
\, .	\end{align}
\end{enumerate} 

\bibliographystyle{JHEP} 
\bibliography{defects}
\end{document}